\documentclass[useAMS,usenatbib]{mn2e}
\usepackage{amsmath}
\usepackage{multicol}
\usepackage{epsfig}
\usepackage{rotating}
\usepackage{color}

\title[M. E. Bell et al]
{Time-domain and spectral properties of pulsars at 154 MHz}
\author[M.E.Bell et al]{M. E. Bell$^{1,2}$\thanks{E-mail:
martin.bell@csiro.au (MEB)}, Tara Murphy$^{2,3}$, S. Johnston$^{1}$,
  D.~L. Kaplan$^{4}$, S. Croft$^{5,6}$, P. Hancock$^{2,7}$, 
\newauthor J.~R.~Callingham$^{1,2,3}$,
A.~Zic$^{3}$, 
D.~Dobie$^{3}$,  
J. K. Swiggum$^{4}$,
A. Rowlinson$^{8,9}$, \and
N. Hurley-Walker$^{7}$ 
A. R. Offringa$^{9}$, 
G.~Bernardi$^{10,11,12}$, 
J.~D.~Bowman$^{13}$, 
F.~Briggs$^{14}$, \and
R.~J.~Cappallo$^{15}$, 
A.~A.~Deshpande$^{16}$, 
B.~M.~Gaensler$^{2,3,17}$, 
L.~J.~Greenhill$^{12}$, \and
B.~J.~Hazelton$^{18}$, 
M.~Johnston-Hollitt$^{19}$, 
C.~J.~Lonsdale$^{15}$, 
S.~R.~McWhirter$^{15}$, \and 
D.~A.~Mitchell$^{1,2}$,  
M.~F.~Morales$^{18}$,  
E.~Morgan$^{5}$, 
D.~Oberoi$^{19}$, 
S.~M.~Ord$^{2,7}$,
T.~Prabu$^{16}$, \and
N.~Udaya~Shankar$^{16}$, 
K.~S.~Srivani$^{16}$, 
R.~Subrahmanyan$^{2,16}$, 
S. J. Tingay$^{2,7}$, \and
R.~B.~Wayth$^{2,7}$, 
R.~L.~Webster$^{2,20}$, 
A.~Williams$^{7}$, 
C.~L.~Williams$^{15}$ \\
\\
$^{1}$CSIRO Astronomy and Space Science, PO Box 76, Epping NSW 1710, Australia\\
$^{2}$ARC Centre of Excellence for All-sky Astrophysics (CAASTRO) \\
$^{3}$Sydney Institute for Astronomy (SIfA), School of Physics, The University of Sydney, NSW 2006, Australia \\
$^{4}$Department of Physics, University of Wisconsin-Milwaukee, 1900 E. Kenwood Boulevard, Milwaukee, WI 53211, USA \\
$^{5}$University of California, Berkeley, Astronomy Dept., 501 Campbell Hall \#3411, Berkeley, CA 94720, USA \\
$^{6}$Eureka Scientific, Inc., 2452 Delmer Street Suite 100, Oakland, CA 94602, USA \\
$^{7}$International Centre for Radio Astronomy Research, Curtin University, Bentley, WA 6845, Australia\\
$^{8}$Anton Pannekoek Institute, University of Amsterdam, Postbus 94249, 1090 GE, Amsterdam, The Netherlands\\
$^{9}$Netherlands Institute for Radio Astronomy (ASTRON), PO Box 2, 7990 AA Dwingeloo, The Netherlands \\
$^{10}$SKA SA, 3rd Floor, The Park, Park Road, Pinelands, 7405, South Africa \\
$^{11}$Department of Physics and Electronics, Rhodes University, PO Box 94, Grahamstown 6140, South Africa\\
$^{12}$Harvard-Smithsonian Center for Astrophysics, Cambridge, MA 02138, USA\\
$^{13}$School of Earth and Space Exploration, Arizona State University, Tempe, AZ 85287, USA\\
$^{14}$Research School of Astronomy and Astrophysics, Australian National University, Canberra, ACT 2611, Australia\\
$^{15}$MIT Haystack Observatory, Westford, MA 01886, USA\\
$^{16}$Raman Research Institute, Bangalore 560080, India\\
$^{17}$Dunlap Institute for Astronomy and Astrophysics, University of Toronto, ON, M5S 3H4, Canada\\
$^{18}$Department of Physics, University of Washington, Seattle, WA 98195, USA\\
$^{19}$School of Chemical \& Physical Sciences, Victoria University of Wellington, PO Box 600, Wellington 6140, New Zealand\\
$^{19}$National Centre for Radio Astrophysics, Tata Institute for Fundamental Research, Pune 411007, India\\
$^{20}$School of Physics, The University of Melbourne, Parkville, VIC 3010, Australia
}

\begin{document}

\date{Accepted xxxx December xx. Received xxxx December xx; in original form xxxx October xx}

\pagerange{\pageref{firstpage}--\pageref{lastpage}} \pubyear{2002}

\maketitle

\label{firstpage}
\begin{abstract}
We present 154~MHz Murchison Widefield Array imaging observations and variability information for 
a sample of pulsars. Over the declination range $-80^{\circ} < \delta < 10^{\circ}$ we detect 17 known pulsars with mean flux density greater than 0.3~Jy. We explore the variability properties of this sample on timescales of minutes to years. For three of these pulsars, PSR~J0953$+$0755, PSR~J0437$-$4715 and PSR~J0630$-$2834 we observe interstellar scintillation and variability on timescales of greater than 2 minutes. One further pulsar, PSR~J0034$-$0721, showed significant variability, the physical origins of which are difficult to determine. The dynamic spectra for PSR~J0953$+$0755 and PSR~J0437$-$4715 show discrete time and frequency structure consistent with diffractive interstellar scintillation and we present the scintillation bandwidth and timescales from these observations. The remaining pulsars within our sample were statistically non-variable. We also explore the spectral properties of this sample and find spectral curvature in pulsars PSR~J0835$-$4510, PSR~J1752$-$2806 and PSR~J0437$-$4715.   
\end{abstract}


\section{Introduction}
The time-varying low-frequency radio sky offers a rich parameter space for exploration. With the advent of low frequency, wide-field, 
and high resolution interferometers e.g. the Murchison Widefield Array (MWA; \citealt{lonsdale_2009}, \citealt{Tingay_MWA}), the 
Low Frequency Array (LOFAR; \citealt{LOFAR}), and the Long Wavelength Array (LWA; \citealt{LWA}) it is now feasible to blindly search vast areas of the sky for transient and variable phenomena. The purpose of such surveys is to explore the physical mechanisms (both intrinsic and extrinsic) driving dynamic behaviour in known and unknown classes of sources. 

In this paper we present time domain measurements of 17 bright pulsars on cadences of minutes, months and years. These measurements
have been made as part of the Murchison Widefield Array Transients Survey (MWATS). MWATS is a time-domain survey covering the declination range $-80^{\circ} < \delta < +10^{\circ}$ at 154~MHz. For this survey, high fidelity wide-field (1000~deg$^{2}$) images were obtained with integration times of just 112 seconds. The science goal of MWATS is to provide a blind low frequency census of transient and variability activity (Bell et al., in prep).

Pulsars are compact stellar remnants that emit regular pulses as they spin, with significant intrinsic variability on timescales shorter than a second.
A small subset of pulsars are known to emit giant radio pulses (e.g. \citealt{Johnston_2011}, \citealt{Tsai_2015}). Giant pulses are typically broadband in nature with a low duty cycle when compared with normal pulses (see \citealt{pulsar_handbook}; \citealt{Oransaye_2015}). Some pulsars show intermitancy on various timescales (e.g. \citealt{int}, \citealt{beta}). For example, nulling i.e. the absence of detectable radio emission for one or more pulse periods, could modulate the long term phase averaged flux density, if the null rate was large (e.g. see \citealt{Deich_1986}). 


However, for most pulsars the average emitted flux density is constant when averaged over suitably long timescales (minutes or longer).  The received flux density can be modulated, though, because of propagation effects such as diffractive and refractive interstellar scintillation \citep{armstrong} that affects pulsars due to their compact sizes ($10^{-3}$~$\mu$arcseconds; \citealt{Lazio_2004}). Diffractive interstellar scintillation is the interference of different paths of a ray, between a source and receiver \citep{Goodman_97}. The different paths arise from small-scale inhomogeneities in the interstellar medium (ISM). Diffractive interstellar scintillation can cause variations on timescales of tens of minutes but is dependent on, for example, dispersion measure, distance, frequency, and pulsar transverse velocity (see \citealt{Rickett_77}; \citealt{Cordes_86}). Refractive interstellar scintillation is caused by large scale electron density irregularities along the line of sight \citep{Bhat_1999} and constitutes a slower and less modulated variation in the pulsar flux density over weeks to months \citep{Sieber}. 

Depending on the cadence of the observations we can explore different variability regimes for different pulsars. Exploring these different regimes has typically been done via high-time resolution observations, rather than imaging (e.g. see \citealt{Stappers_2011}; \citealt{bhat_2014}; \citealt{Tsai_2015}; \citealt{Kondratiev_2015}). Long term studies have specifically aimed at exploring the effects of the ISM. For example, \cite{gupta_93} present daily phase averaged flux densities of nine pulsars over a duration of 400 days. For the majority of pulsars in their sample the flux density changes were consistent with those predicted by refractive interstellar scintillation (also see \citealt{Kaspi_92}; \citealt{stinebring_2000} and \citealt{zhou_2003}).  

Imaging observations can offer an alternative and convenient way of studying and possibly even discovering pulsars (e.g. \citealt{Backer_82}; \citealt{kaplan_98}). With the increased survey speed of next generation wide-field instruments, much of this information comes for free. In this paper we present a time domain survey of a sample of 17 known pulsars. This survey allows us to probe the short and long term effects of the ISM on pulsar flux densities at low frequencies, and more generally the variability properties of this sample. In addition we evaluate the ability for the MWA to study pulsars via imaging observations and its applications to future surveys.   

In Section 2 of this paper we present the observing strategy and pulsar sample selection. 
In Section 3 we discuss the data reduction strategy and variability statistics used to characterise the sample. 
In Section 4 we present the results of our analysis focusing on the pulsars that showed significant variability. 
In Section 5 we discuss our results and explore what might be achieved with similar but deeper surveys using image plane techniques. 

\section{Observing strategy and pulsar sample selection}
Data collection for this survey began in 2013 July and ended in 2015 July. The observing cadence was approximately one night per month, and on each night we typically observed for 10 hours. Observations were conducted at a centre frequency of 154~MHz with an observing bandwidth of 30.72~MHz. A channel bandwidth of 40~KHz and a correlator integration time of either 0.5~s or 2~s were used for these observations. The correlator integration time was increased to 2~s in later observations to reduce data rates.  

We used a drift scanning strategy to cover a large sky area each night. Utilising night-time seasonal sky rotation allows for sampling the entire hemisphere over one year. A given pulsar takes approximately one hour to drift through the primary beam (FWHM of 24.4$^{\circ}$ at 154 MHz).  The observing strategy was to cycle through three different pointings along the meridian at $\delta = -55^{\circ}$, $-26^{\circ}$ (zenith) and +1.6$^{\circ}$. These declination strips overlap giving complete sky coverage between $+10^{\circ}$ and $-80^{\circ}$. A 112 second snapshot observation was obtained at each of these declinations in turn for the duration of the observing run.
Due to the observing strategy, for a given declination, a four minute gap occurs between observations. An additional eight seconds are required to update the correlator configuration for a new pointing.  A summary of the observing specifications are given in Table~\ref{observations}. 

\begin{table}
\centering
\caption{Properties of observations. Typical noise values are quoted in the extragalactic direction for $b>10^{\circ}$ and in the galactic direction for $b<10^{\circ}$.}
\begin{tabular}{|c|c|}
\hline
Property & Value \\
\hline
Integration time per snapshot & 112 seconds \\
Number of snapshots per pulsar & 55$-$159 \\
Cadence & minutes, months and years  \\ 
Image size (pixels) & 3072 $\times$ 3072 \\
Frequency & 154 MHz \\
Bandwidth & 30.72 MHz \\
Channel bandwidth & 40~KHz \\
Pixel diameter & $0.75^{\prime}$ \\
Resolution at 154 MHz & $2.4^{\prime}$ \\
Briggs weighting & $-$1 \\
UV range (k$\lambda$) & $>0.03$ k$\lambda$ \\
Declinations & $+$1.6$^{\circ}$, $-$26$^{\circ}$, $-$55$^{\circ}$ \\
Typical image noise\\ 
(extragalactic pointing) & 20~mJy\\
Typical image noise\\ 
(galactic pointing) & 100~mJy\\
\hline
\label{observations}
\end{tabular}
\end{table}

Two data products were generated from this survey: (1) single snapshot images, used to generate the light curves of the pulsars (discussed below); and (2) mosaiced monthly images formed from all snapshots for a given declination. These images were used for the initial identification of the pulsars in our sample.   

We used the Australia Telescope National Facility (ATNF) pulsar database\footnote{http://www.atnf.csiro.au/people/pulsar/psrcat/} (version 1.54; date accessed 2015-05-01) to determine positions of known pulsars in our survey region. There were 2297 known pulsars in our survey region of declination $<+10^{\circ}$. We searched for detections at the positions of each pulsar in our monthly mosaiced images. If a detection was made the statistics were recorded e.g signal-to-noise ratio, flux density etc. Of a total 2297 pulsars that were within our sky area over 100 were detected above the 3$\sigma$ noise level. For this analysis we focused on extracting variability information, so we concentrated on bright, well detected pulsars that had adequate signal-to-noise ratio ($>8\sigma$) in the monthly mosiaced images. This restricted our final sample of pulsars to 17 (see Table~\ref{pulsar_table} for details). A more complete analysis of all pulsar detections will be presented in future work.    

 \section{Data reduction}
 \subsection{Phase calibration, flagging, imaging, and self-calibration}
Phase calibration was performed as follows.
 \label{cal_section}
 A snapshot observation (with integration time 112~seconds) of a well modelled bright source was obtained for phase calibration 
purposes as a function of declination strip and observing run. Model images of these calibrator 
sources were extracted from the Sydney University Molonglo Sky Survey (SUMSS; \citealt{SUMSS}) or the VLA Low-frequency Sky Survey  (VLSS; \citealt{VLSS}). 
The model image of the calibrator source was inverse-Fourier transformed to generate a set of model visibilities.  A single 
time-independent, frequency-dependent amplitude and phase calibration solution was derived from 
this model with respect to the calibrator observation visibilities. These gain solutions were then applied to the appropriate target visibilities (discussed below). We will discuss flux density scale corrections in Section \ref{flux_scale}.

For each of the snapshot target observations we performed the following processing: 
\begin{itemize}
\item Data were flagged for radio frequency interference using the {\sc aoflagger} algorithm (\citealt{offringa_2012}) and converted into {\sc casa} measurement set format using the MWA preprocessing pipeline {\sc cotter}. Approximately 1\% of the visibilities were removed at this stage, see \citealt{Andre_2015} for a thorough discussion; 
\item Phase and amplitude calibration solutions were applied to the visibilities (as discussed above); 
\item The visibilities were deconvolved and {\sc clean}ed with 2000 iterations using the 
{\sc wsclean} algorithm \citep{offringa_2014}. An RMS noise measurement was taken from the images to 
ascertain an appropriate {\sc clean} threshold for post self-calibration imaging; 
\item The {\sc clean} component model was inverse Fourier transformed for self-calibration purposes. 
A new set of phase and amplitude calibration solutions were derived from this model and applied to the 
data; 
\item The visibilities were then deconvolved and {\sc clean}ed to a cutoff of three times the RMS derived from the pre self-calibration image. An image size of 3072 $\times$ 3072 with pixel diameter 0.75$^{\prime}$ and robust parameter of $-1$ was used; 
\item A primary beam correction was applied to create Stokes I images. See \cite{offringa_2014} for further details. 
\end{itemize}

As discussed above, two different data products were generated from the data reduction. First, we reduced a smaller subset of the total data covering approximately one year and all of our survey area. For a given night and declination strip all snapshot images were mosaiced together. We used these mosaics to construct our initial sample of detected pulsars. Second, we reduced all available images for our sample of 17 pulsars to produce complete light-curves. For each detected pulsar location we obtained all MWATS observations that were within a radius of 12$^{\circ}$. These observations were then reduced and imaged as discussed above. We aimed to image pulsars within 12$^{\circ}$ of the pointing centre to reduce the effects of uncertain primary beam correction (discussed further below). This is also to mitigate against the drop-off in sensitivity towards the edge of the beam. Two of the pulsars (PSR~J$0034-0721$ and PSR~J1456$-$6843) were located greater than 12$^{\circ}$ from our pointing centre but we include them in this analysis. This is because they are bright with low dispersion measures and as such we predicted that we might be able to detect variability.   

\subsection{Flux density scale correction} 
\label{flux_scale}
\subsubsection{Relative flux scale}
We calibrated the relative flux density scale of each snapshot image. This calibration consisted of comparing the flux density of unresolved sources detected within each image, to the flux density of sources from the SUMSS or VLSS catalogs. The SUMSS catalog was used for images south of $-30^{\circ}$ declination, whilst the VLSS was used for sources north of this declination. For each snapshot image we calculate the mean ratio of the MWA sources to that expected from either of the reference catalogs. Since the SUMSS, VLSS, and MWATS surveys are all at different frequencies (843~MHz, 74~MHz, and 154~MHz respectively), we scaled the reference catalog flux densities to the MWATS frequency using a spectral index of $\alpha=-0.8$ \citep{Lane_2014}. The mean flux density ratio $f_{g}$ was then used to correct all the MWATS flux densities to be in line with the SUMSS or VLSS flux densities. 

This method bootstraps the flux density scale of an ensemble of unresolved sources (in the MWA images) rather than from a single source, and ensures an internally consistent flux scale. Typically between 150$-$500 crossmatched sources are used for this calculation. For sources that are not expected to be variable, we see an epoch-to-epoch flux density variation of $2-5$\% (calculated using approximately 1000 sources per pulsar field). We take this to be the accuracy of our relative flux density calibration. 

\subsubsection{Absolute flux scale}
The method described above achieves a good relative flux density scale between epochs; it does not, however, guarantee that the absolute flux density scale is well calibrated with respect to other radio catalogues. 
It is an area of active research to adequately constrain the low frequency flux density scale in the Southern Hemisphere (e.g. see \citealt{Joe}, \citealt{MWACS} and \citealt{GLEAM}). 
Noting the absolute flux density scale is uncertain we find the relative flux density scale correction between images to be sufficient to achieve the goals of this work. 

\subsection{Light curve extraction} 
The light curves of the pulsars were extracted using a forced fit algorithm implemented in the {\sc aegean} (version 1.9.5) source finding software package \citep{aegean}.The right ascension and declination of each of the pulsars were fitted in the respective images to return the flux density values. The beam properties recorded in the image headers were used to constrain the Gaussian fit. We used the peak flux density reported by {\sc aegean} for all subsequent analyses. We also fitted two neighbouring unresolved sources that had a signal-to-noise ratio of above eight. The modulation indexes of these neighbouring sources were used to ascertain errors on the flux stability of the instrument. This will be discussed further in Section \ref{error}.

Due to the small angular sizes of the pulsars they should be unresolved at the MWA resolution. We visually inspected a region within a radius $5^{\prime}$ surrounding the pulsar positions for bright extended Galactic plane emission. Pulsars embedded in these complex regions were removed from our final sample. Extended emission can cause complications in obtaining adequate and stable measurements of flux density.   

\subsection{Variability statistics}
\label{var_stats_section}
For each pulsar light-curve we calculated the reduced $\chi^{2}_{r}$ statistic. We used the assumption that the light-curve of a given pulsar was non-variable and the weighted mean of the flux density measurements was used as a model for the test. The reduced $\chi^{2}_{r}$ statistic is defined as:
 
\begin{equation}
\chi_{r}^{2} = \frac{1}{n-1}  \sum_{i=1}^{n} \frac{(S_{i} - \tilde{S})^{2}}{\sigma_{i}^{2}},
\label{chisquared}
\end{equation}

\noindent where $S_{i}$ is the i$th$ flux density measurement with variance $\sigma^{2}_{i}$ and $n$ is the total number of epochs. The weighted mean flux density, $\tilde{S}$, is defined as 

\begin{equation}
\tilde{S} = \sum_{i=1}^{n} \left( \frac{ S_{i}}{\sigma_{i}^{2}} \right) /  \sum_{i=1}^{n} \left( \frac{1}{\sigma_{i}^{2}} \right).
\label{w_mean}
\end{equation}

We also calculate the modulation index which is defined as:

\begin{equation}
M = 100 \times( \sigma / \overline{S}),
\end{equation}

\noindent where $\sigma$ is the standard deviation of the flux density measurements and $\overline{S}$ is the arithmetic mean (not the weighted mean).  

\subsection{Error analysis}
\label{error}
The errors reported by {\sc aegean} give a good characterisation of the error in fitting a Gaussian to a point source in a single image. There are a number of other sources of error in our flux density measurements:
\begin{itemize}

\item Primary beam errors: The precise primary beam response of the MWA is difficult to model with increasing distance away from the pointing centre \citep{Sutinjo_2015}.
To reduce this effect we limit our analysis to within 12$^{\circ}$ of the pointing centre where this error is estimated to be around 5\% (see \citealt{cleo_pb}). We apply this restriction to 15 of the pulsars in our sample. For two of the pulsars, PSR~J0034$-$0721 and PSR~J1400$-$6325, this was impractical and we allowed measurements within 15$^{\circ}$ of the pointing centre; 

\item Flux density scale correction errors: The flux density scale correction discussed in Section \ref{flux_scale} is not robust to problem images e.g. those containing bright diffuse Galactic emission in the sidelobes. Images with extreme flux density scale corrections $0.1<f_{g}<1.9$ were removed from this analysis. Images requiring extreme flux density scale corrections were often of poor quality. The resulting light curves obtained from using those images typically contained excess non-physical variability, which was clearly correlated with the extreme flux density scale corrections. 

The range of corrections ($f_{g}$) we accept represents the different calibrator models that we have used for phase calibration. Note, the initial flux density scale of these calibrators was never intended for absolute flux calibration (hence the need for a robust flux density scale correction). One of the calibrators we used required flux scale corrections $f_{g}\sim0.15$ to bring the images onto a common flux scale. This resulted in a skewing of the acceptable flux scale corrections we used in the final light-curves;   
  
\item Ionospheric: Excited geomagnetic conditions can distort the location of background radio sources (e.g. see \citealt{Cleo}) which can in turn affect the accuracy of the flux scale correction and source fitting algorithms. For example, when a number of bright MWA sources were incorrectly crossmatched with SUMSS counterparts, causing incorrect flux scale correction factors ($f_{g}$) and thus flux scale errors. Observations taken during heightened ionospheric activity, which had large positional offsets were removed from this analysis. This accounted for approximately $1\%$ of the total data.   
\end{itemize}

All of the effects described above are difficult to separate out into individual time and position dependent error terms. We therefore boot-strapped our errors from two neighbouring sources (to the given pulsar) of similar flux, under the assumption that they were non-variable. For two sources we measured the averaged modulation index $\overline{M}$ and added this in quadrature with the {\sc aegean} Gaussian errors $e^{2}_{fit}$ and source flux density $S_{i}$ as follows: 

\begin{equation}
e_{i} = \sqrt{e_{fit}^{2}+\left( S_{i} \times \overline{M}\right)^{2}}. 
\end{equation}
  
\noindent $e_{i}$ is the adjusted error on an individual pulsar flux density measurement $S_{i}$. 
By bootstrapping the errors in this way we set the minimum variability that we are capable of 
detecting to that of the neighbouring sources. These are the errors used in the variability 
statistics described in Section \ref{var_stats_section}.

\begin{figure*}
\centering
\includegraphics[scale=0.75]{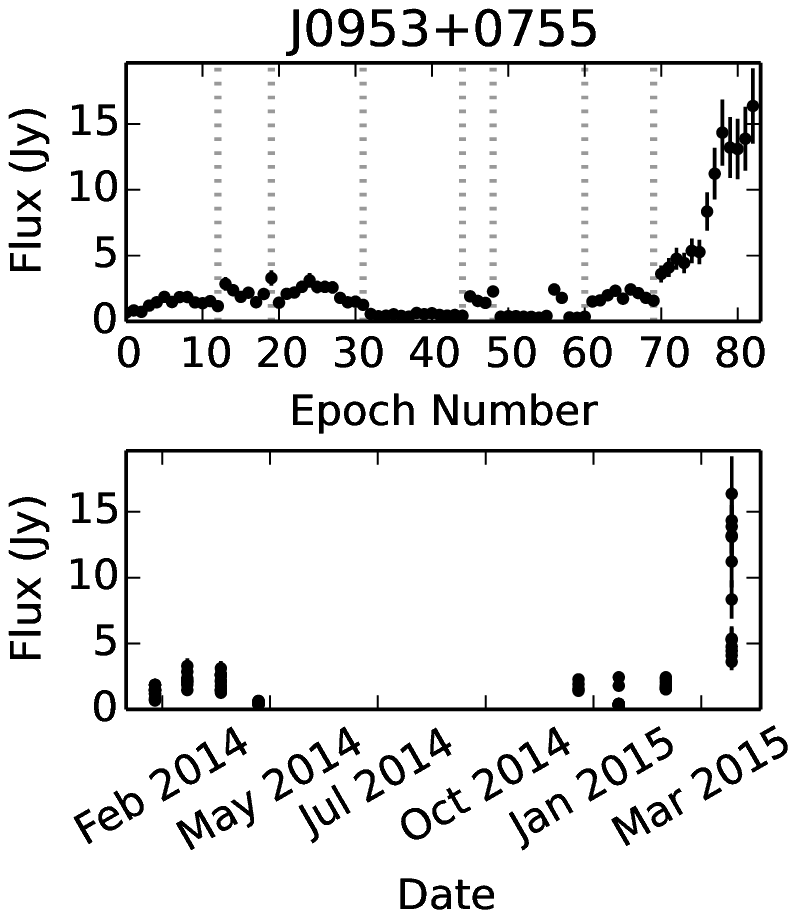}
\includegraphics[scale=0.75]{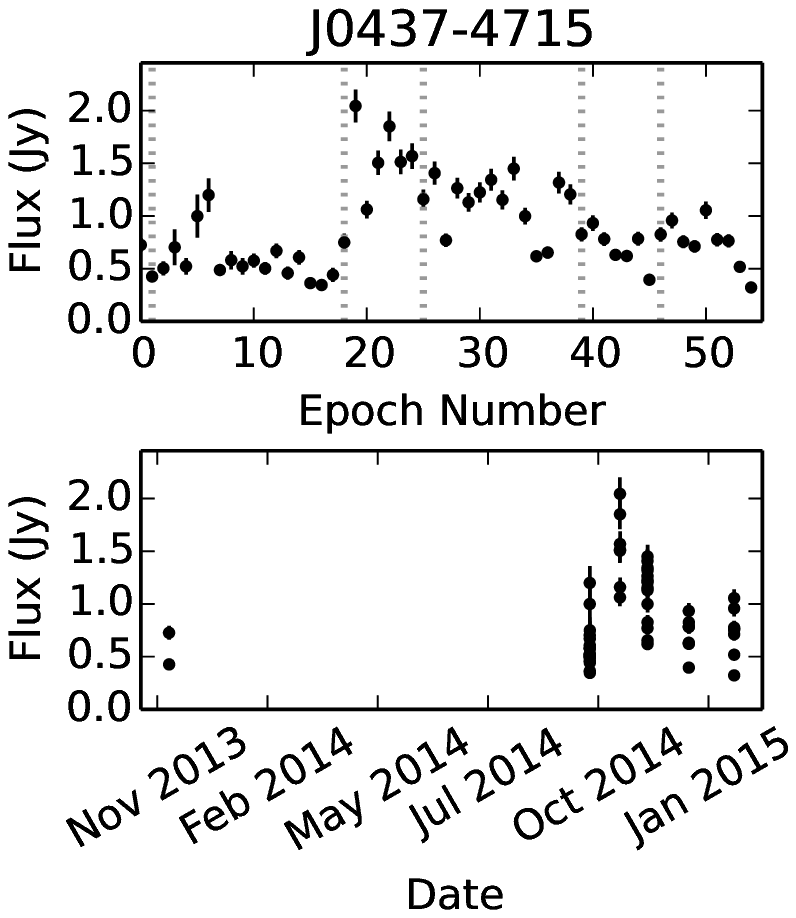}
\includegraphics[scale=0.75]{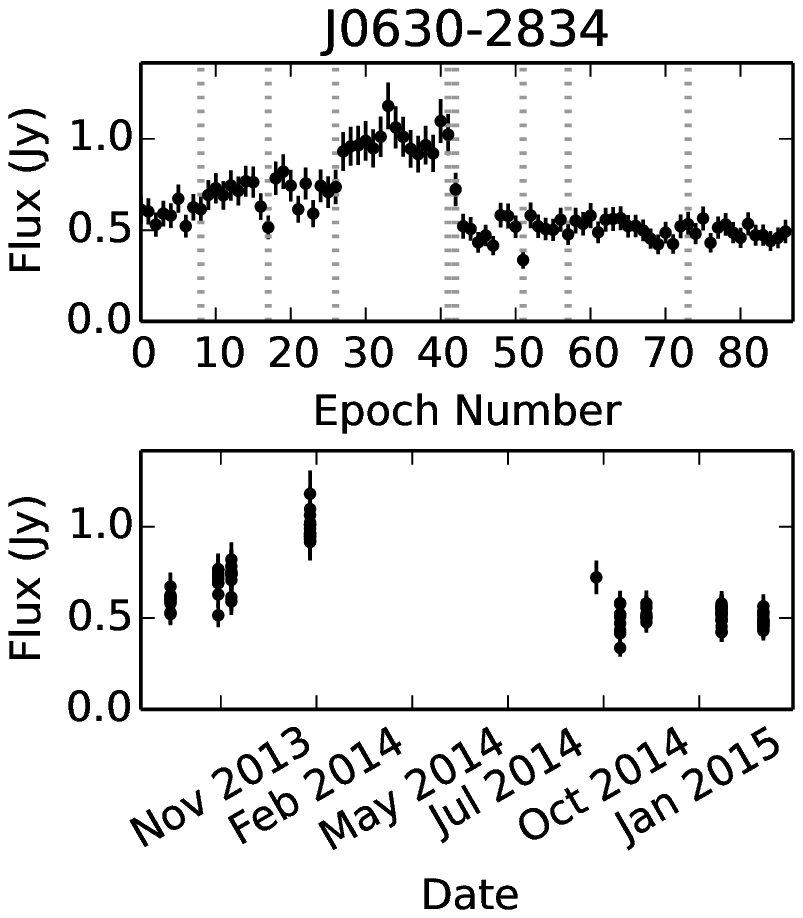}
\includegraphics[scale=0.75]{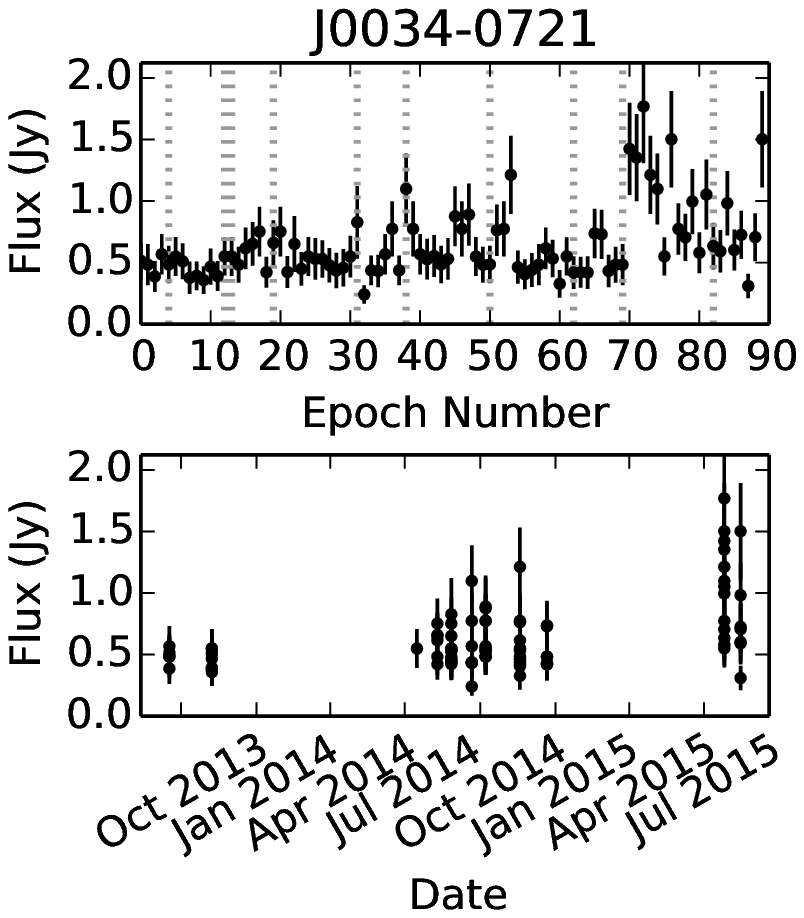}
\includegraphics[scale=0.75]{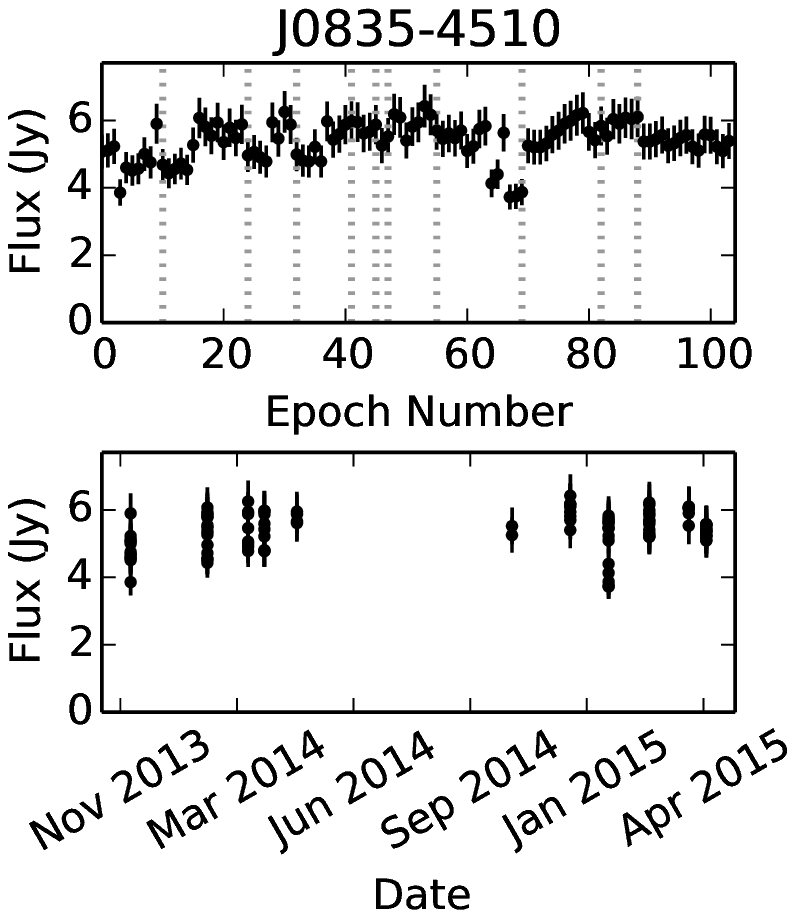}
\includegraphics[scale=0.75]{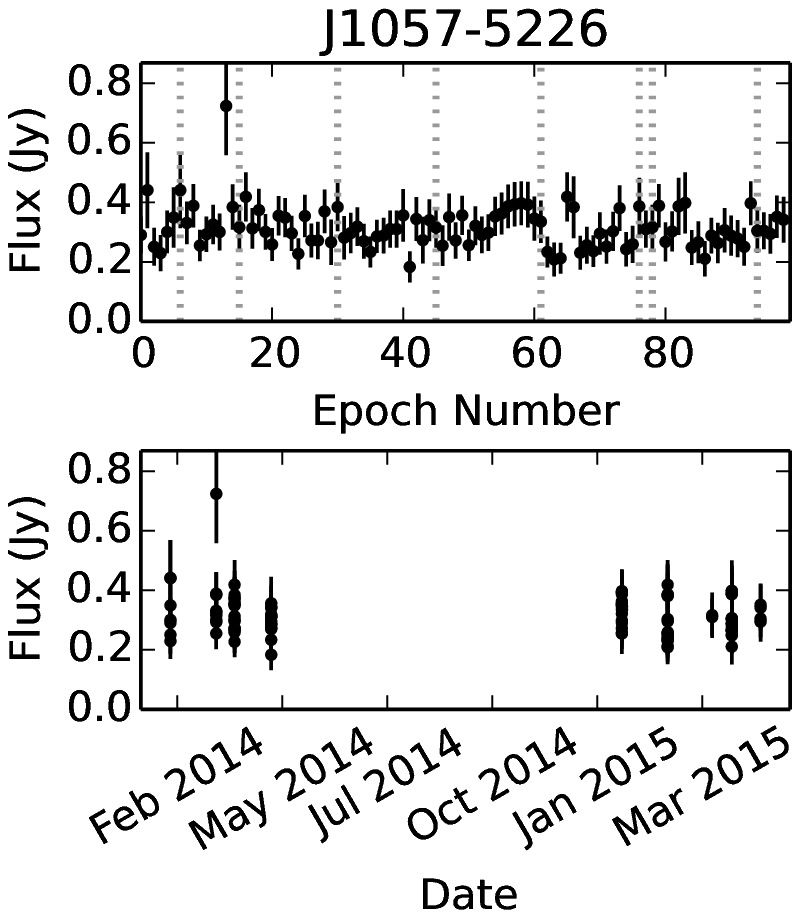}
\caption{Light-curves for nine pulsars detected at high signal-to-noise ratio with the MWA. The top panel of each subplot shows the flux density as a function of sequential epoch number. The bottom panel shows the flux density as a function of date. The dashed grey line denotes an epoch number at which the time difference to the previous observation was greater than 8 days.}
\label{pulsar_lightcurves}
\end{figure*}

\begin{figure*}
\centering
\includegraphics[scale=0.75]{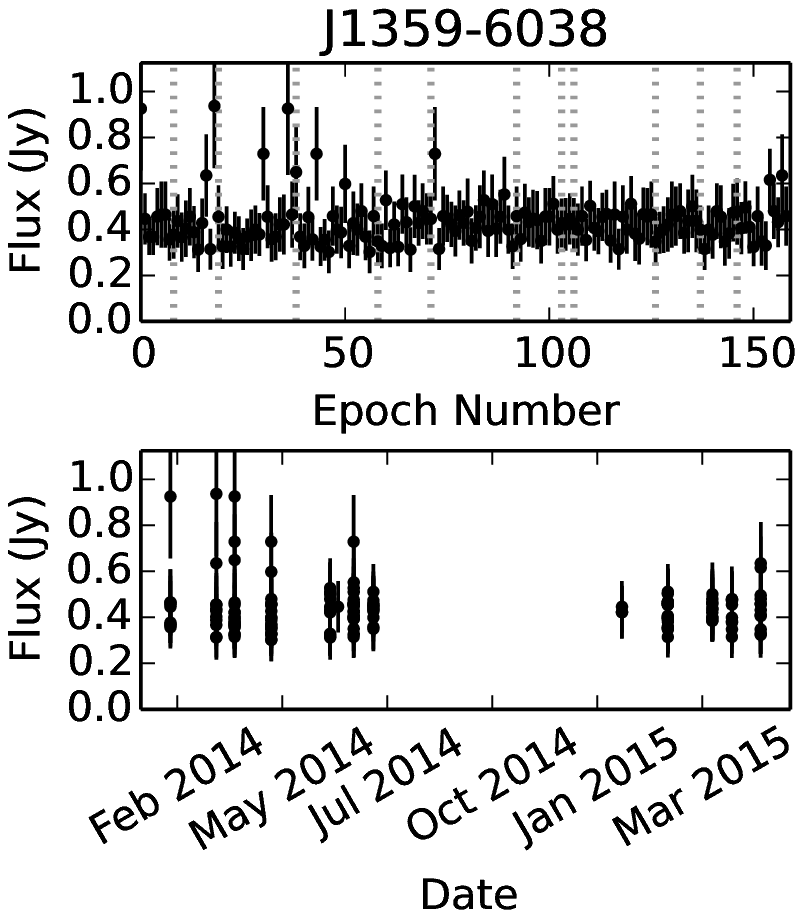}
\includegraphics[scale=0.75]{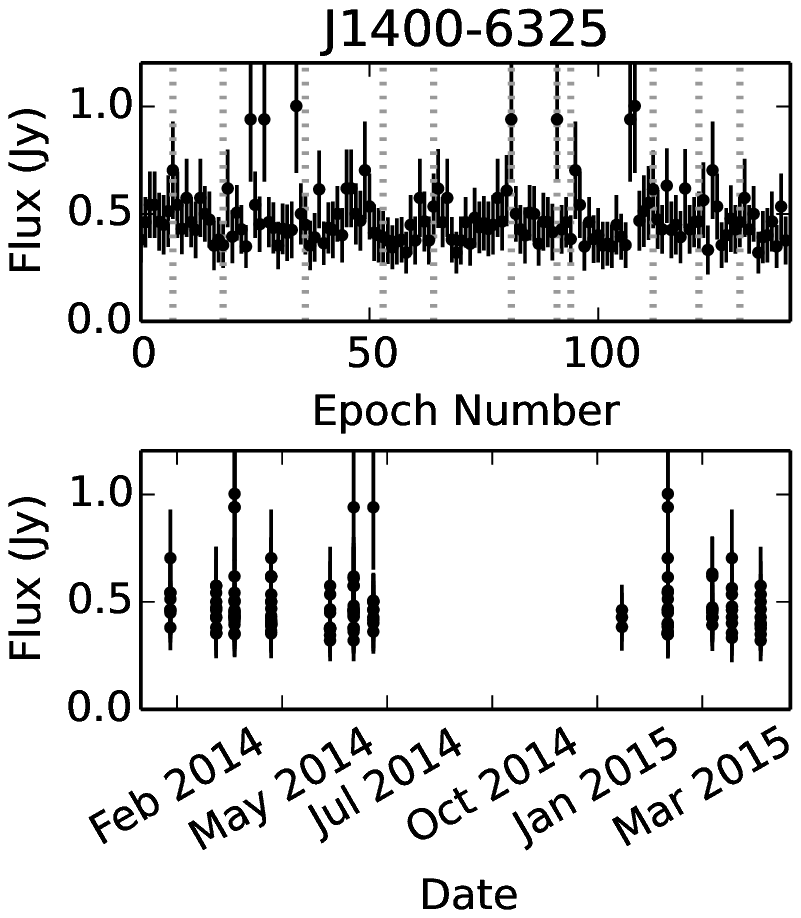}
\includegraphics[scale=0.75]{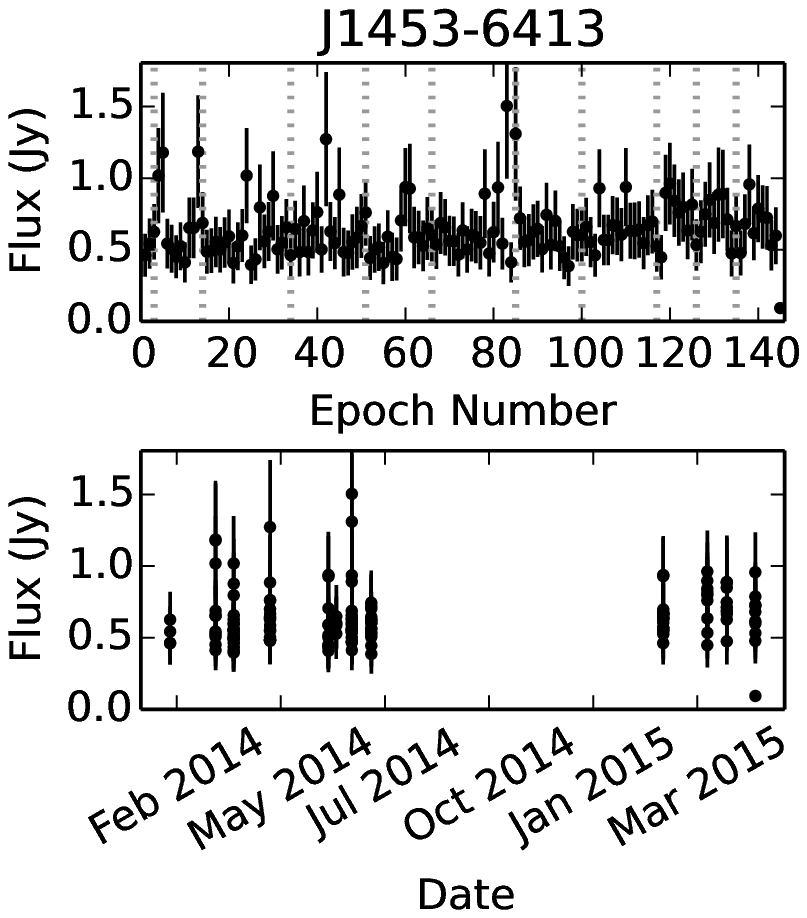}
\includegraphics[scale=0.75]{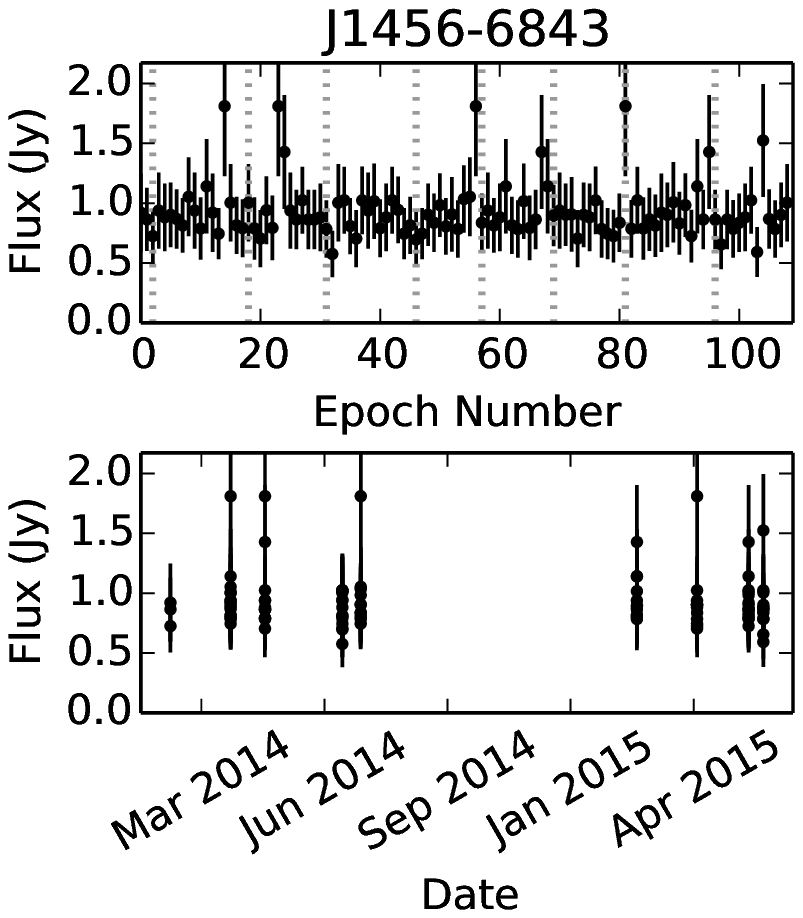}
\includegraphics[scale=0.75]{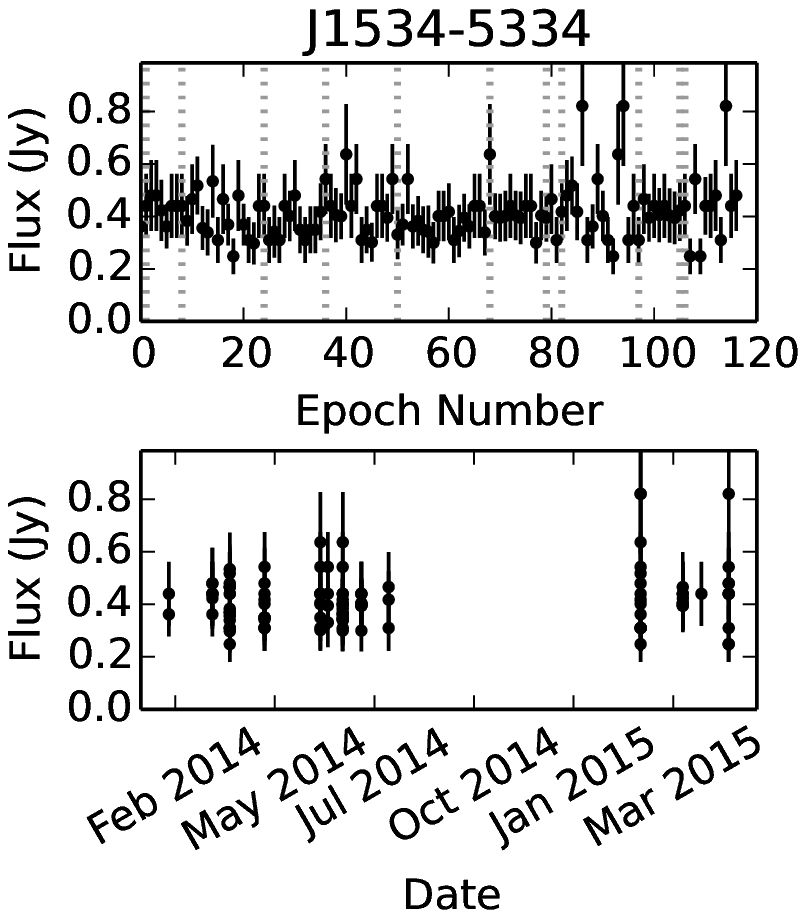}
\includegraphics[scale=0.75]{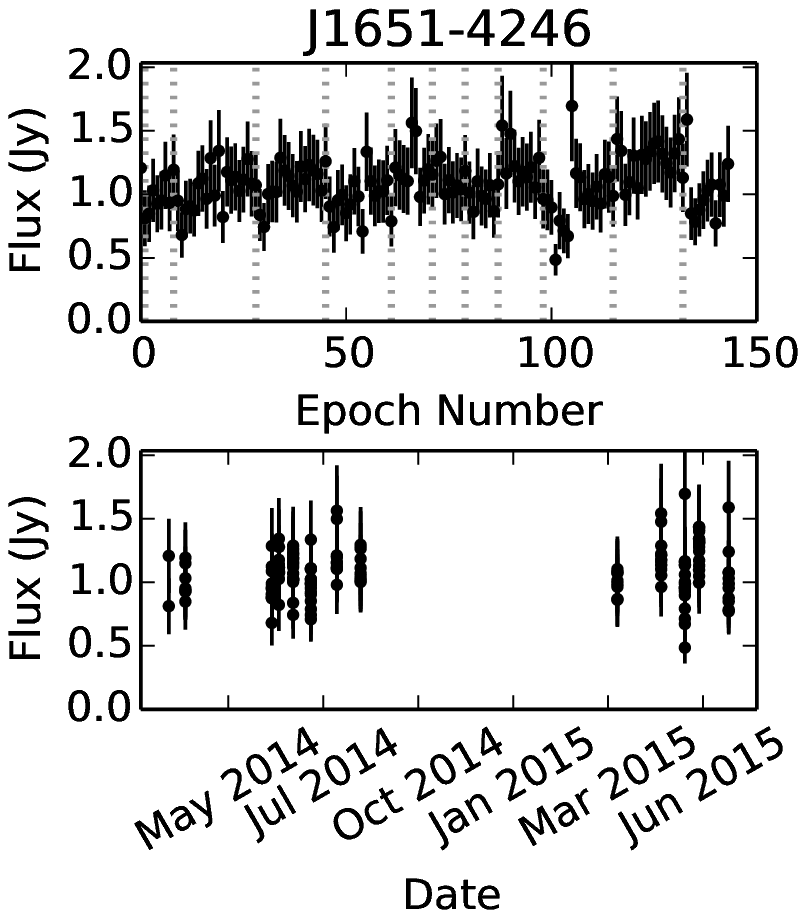}
\caption{Continued: Light-curves for eight pulsars detected at high signal-to-noise ratio with the MWA. Details as in Figure~\ref{pulsar_lightcurves}.}
\label{pulsar_lightcurves2}
\end{figure*}

\begin{figure*}
\centering
\includegraphics[scale=0.75]{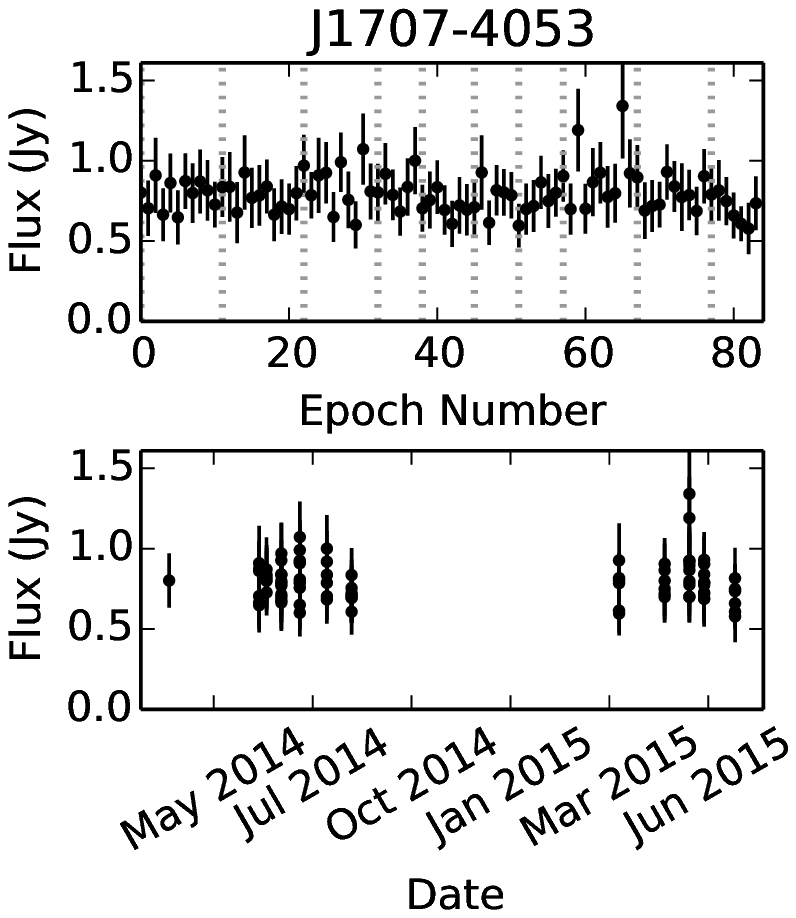}
\includegraphics[scale=0.75]{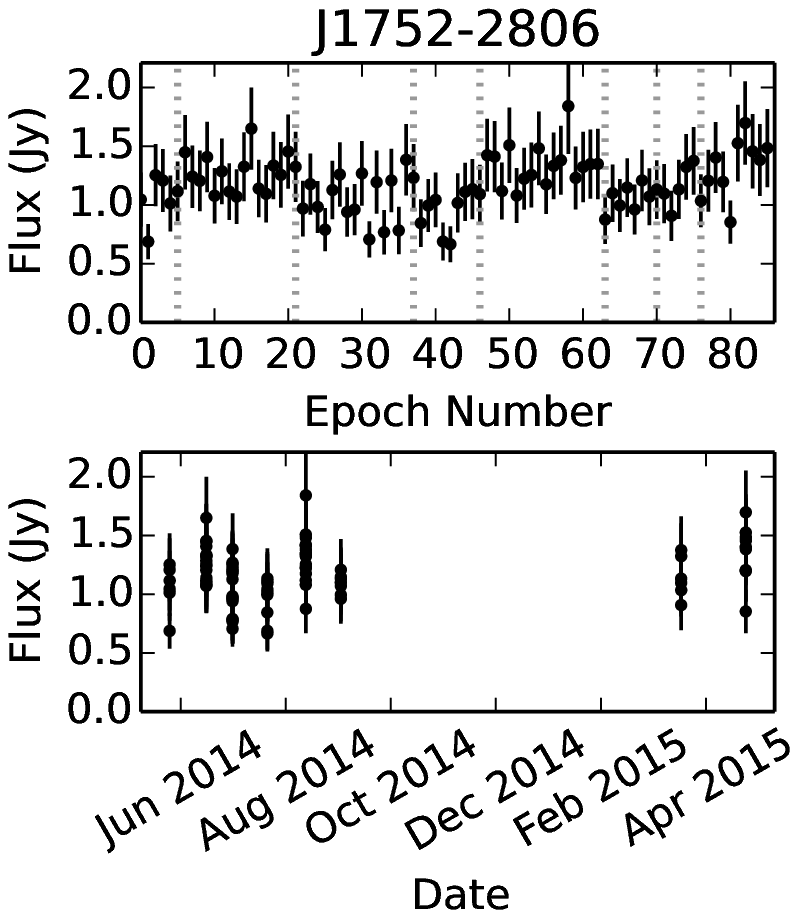}
\includegraphics[scale=0.75]{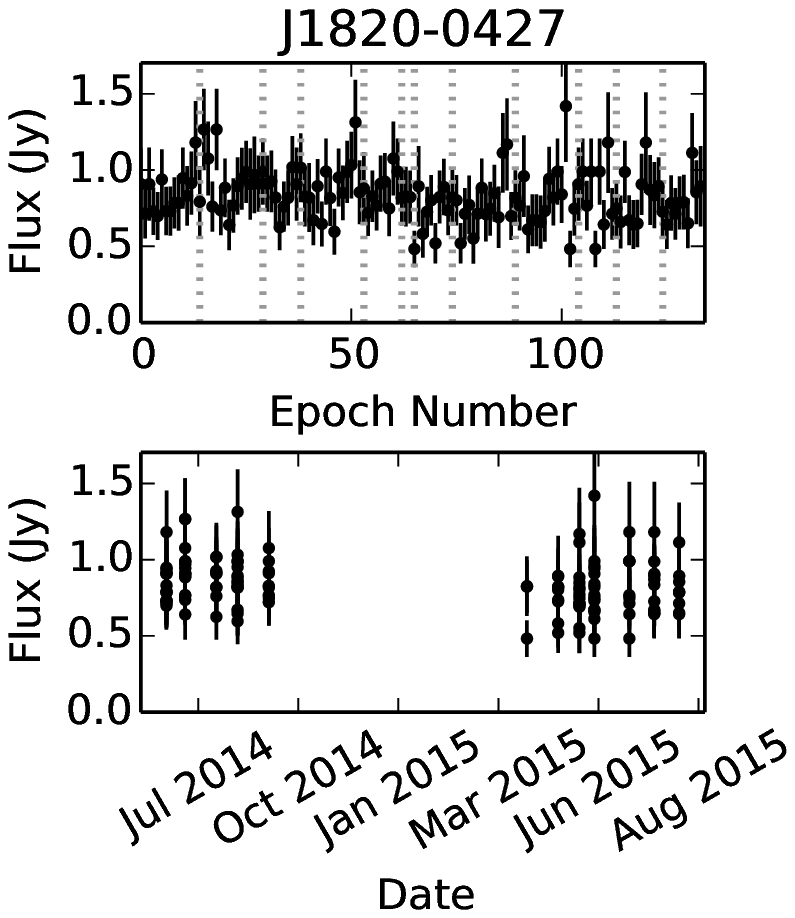}
\includegraphics[scale=0.75]{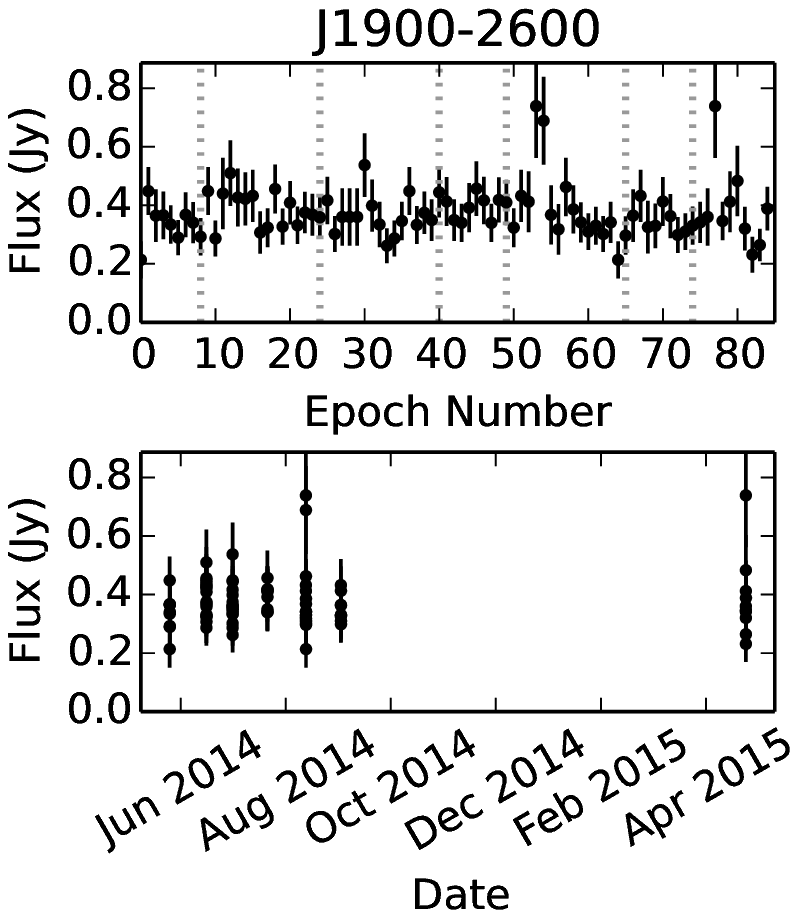}
\includegraphics[scale=0.75]{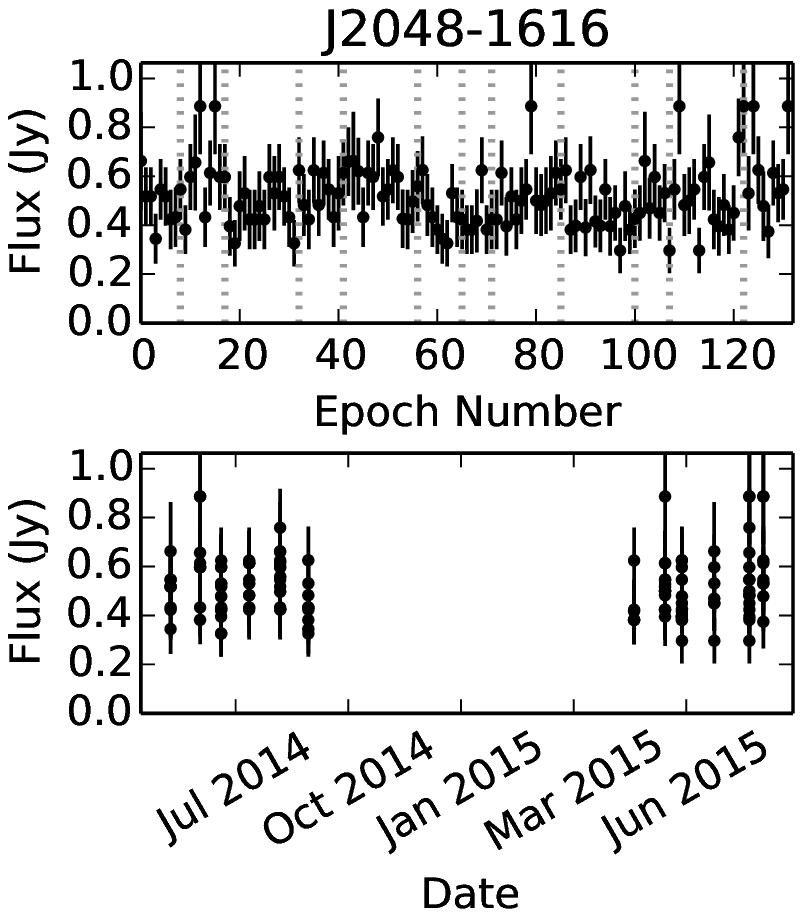}
\caption{Continued: Light-curves for eight pulsars detected at high signal-to-noise ratio with the MWA. Details as in Figure~\ref{pulsar_lightcurves}.}
\label{pulsar_lightcurves3}
\end{figure*}

\section{Results}
\label{results}
We consider a source to be statistically variable if $\chi_{r}^{2}>2.0$. We made a low cut on the minimum $\chi^{2}_{r}$ used to define variability as we have been conservative with our error propagation. Of the 17 pulsars four showed significant variability and we discuss these below. A summary of our results is given in Table~\ref{pulsar_table} and pulsar light-curves are shown in Figures \ref{pulsar_lightcurves}, \ref{pulsar_lightcurves2} and \ref{pulsar_lightcurves3}. 

\begin{table*}
\centering
\caption{Variability of pulsars in this sample. Above the horizontal line is the pulsars that showed significant variability sorted by $\chi^{2}_{r}$. The pulsars below the line remain non-variable and are sorted by right ascension. The column labelled $M$ indicates the modulation index of the pulsar, the $\overline{M}$ column indicates the average modulation index of two nearby sources. The minimum, maximum and average flux densities of the pulsars are denoted by $S_{min}$, $S_{max}$ and $\overline{S}$, respectively. The total number of observations is denoted by $N$. Pulsars located $>12^{\circ}$ from the pointing centre of the observations are marked with a $\dagger$ symbol.}
\begin{tabular}{|l|l|r|r|r|r|r|r|r|r|r|}
\hline
\multicolumn{1}{l}{Pulsar name} & \multicolumn{1}{l}{B name} & \multicolumn{1}{r}{DM (cm$^{-3}$ pc)} & \multicolumn{1}{r}{$M$ (\%)} & $S_{min}$ (Jy) & $S_{max}$ (Jy) & $\overline{S}$ (Jy) &$N$ & $\chi^{2}_{r}$ & $\overline{M}$ (\%)  \\
\hline
PSR~J0953$+$0755 & B0950$+$08 & 2.95 & 131.3  &  0.27  &  16.4 & 2.6 $\pm 0.8$ &  83  &  182.1  &   17.4 \\
PSR~J0437$-$4715 & $-$ & 2.65 & 44.9  &  0.32  &  2.0 & 0.87 $\pm$ 0.3  &  55  &  28.1  &   7.5 \\
PSR~J0630$-$2834 & B0628$-$28 & 34.5  & 30.0  &  0.33  &  1.18 & 0.64 $\pm$ 0.2  &  87  &  5.8  &   10.5 \\
PSR~J0034$-$0721$^{\dagger}$ &  B0031$-$07 & 11.4 & 45.0  &  0.24  &  1.8 & $0.64 \pm 0.2$  &  90  &  2.0  &   25.7  \\
\hline
PSR~J0835$-$4510 & B0833$-$45 & 68.0  & 10.7  &  3.7  &  6.4 & 5.4 $\pm 1.6$  &  104  &  1.5  &   9.8 \\
PSR~J1057$-$5226 & B1055$-$52 & 30.1  & 22.2  &  0.18  &  0.72 & 0.31 $\pm$ 0.1  &  99  &  0.8  &   15.5 \\ 
PSR~J1359$-$6038 & B1356$-$60 & 293.7 & 24.0  &  0.30  &  0.93 & 0.43 $\pm$ 0.1  &  159  &  0.50  &   18.6 \\
PSR~J1400$-$6325$^{\dagger}$ & $-$ & 563.0 & 28.5  &  0.32  &  1.00 & 0.48 $\pm$ 0.1 &  142  &  0.61  &   20.8 \\
PSR~J1453$-$6413 & B1449$-$64 & 70.1 & 27.9  &  0.39  &  1.31 & 0.63 $\pm$ 0.2  &  134  &  0.61  &   27.2 \\
PSR~J1456$-$6843 & B1451$-$68 & 8.6 & 24.8  &  0.58  &  1.8 & $0.93 \pm 0.3$  &  109  &  0.4  &   24.3 \\
PSR~J1534$-$5334 & B1530$-$53 & 24.8 & 24.5  &  0.25  &  0.82 & 0.42 $\pm$ 0.1  &  117  &  0.75  &   15.1 \\
PSR~J1651$-$4246 & B1648$-$42 & 482.0 & 18.2  &  0.48  &  1.70 & 1.08 $\pm$ 0.3  &  144  &  0.75  &   21.0 \\
PSR~J1707$-$4053 & B1703$-$40 & 360.0  & 16.0  &  0.58  &  1.3 & 0.80 $\pm$ 0.2  &  84  &  0.45  &   14.1 \\
PSR~J1752$-$2806 & B1749$-$28 & 50.4  & 20.0  &  0.67  &  1.84 & 1.17 $\pm$ 0.4  &  86  &  1.0  &   20.0 \\
PSR~J1820$-$0427 & B1818$-$04 & 84.4 & 20.2  &  0.48  &  1.41 & 0.83 $\pm$ 0.2 &  134  &  0.81  &   16.6\\
PSR~J1900$-$2600 & B1857$-$26 & 38.0  & 24.0  &  0.21  &  0.73 & 0.37 $\pm$ 0.1  &  85  &  0.93  &   16.2  \\
PSR~J2048$-$1616 & B2045$-$16 &11.5 & 25.0  &  0.29  &  0.89 & 0.52 $\pm$ 0.1 &  132  &  0.93  &   18.5 \\
\hline
\label{pulsar_table}
\end{tabular}
\end{table*}

\subsection{PSR J0953$+$0755 (B0950+08)}
\label{J0953}
We detected significant variability in PSR~J0953$+$0755, with a modulation index over all epochs of $M=131.3\%$ and a $\chi^{2}_{r} = 182.1$ (see Figure \ref{pulsar_lightcurves}). On one of the nights of observing (2015-04-14), extreme variability was detected. For approximately one hour the flux density of the pulsar increased and peaked at $S_{max}=16.4$~Jy (see Table~\ref{pulsar_table}). PSR~J0953$+$0755 \citep{Pil_68} has a low dispersion measure (DM) of 2.95 cm$^{-3}$~pc and spin period of 0.25~s \citep{Hobbs_2004}.  
This pulsar is known to scintillate at low frequencies, for example, \cite{P_C_Nat_1992} report 
observations consistent with diffractive scintillation. 

\begin{figure*}
\centering
\includegraphics[scale=0.57]{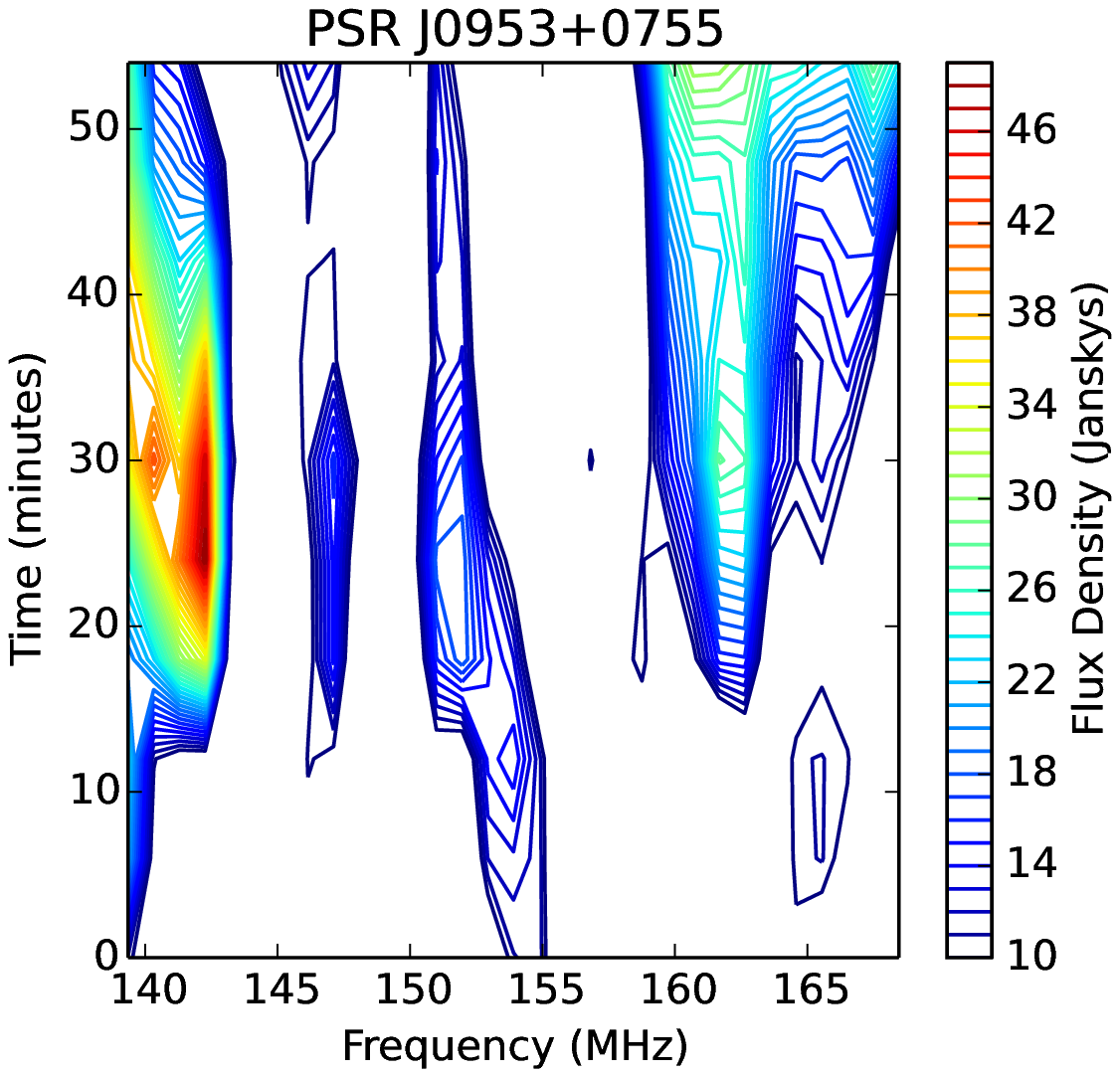}
\includegraphics[scale=0.57]{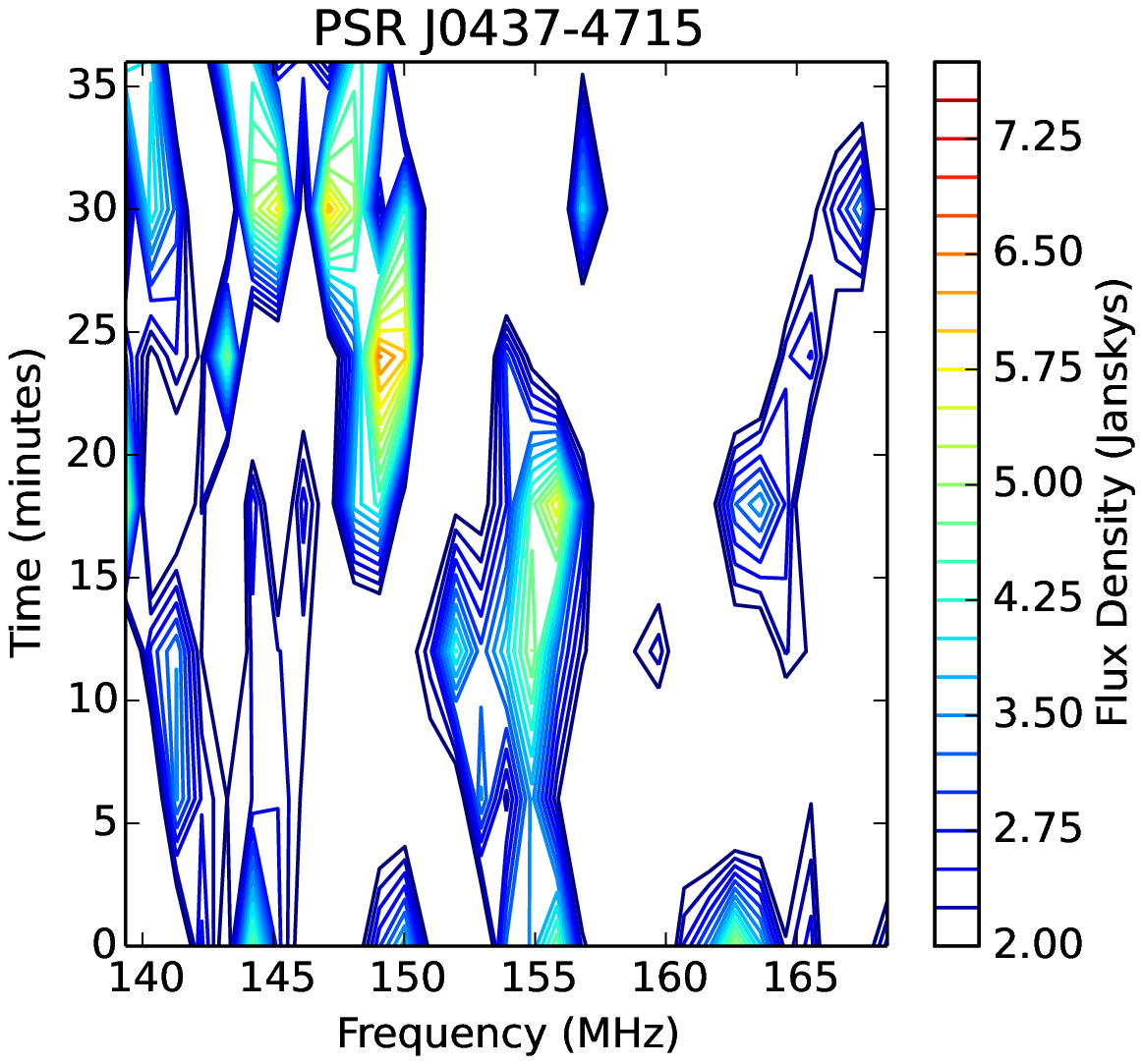} \\
\includegraphics[scale=0.57]{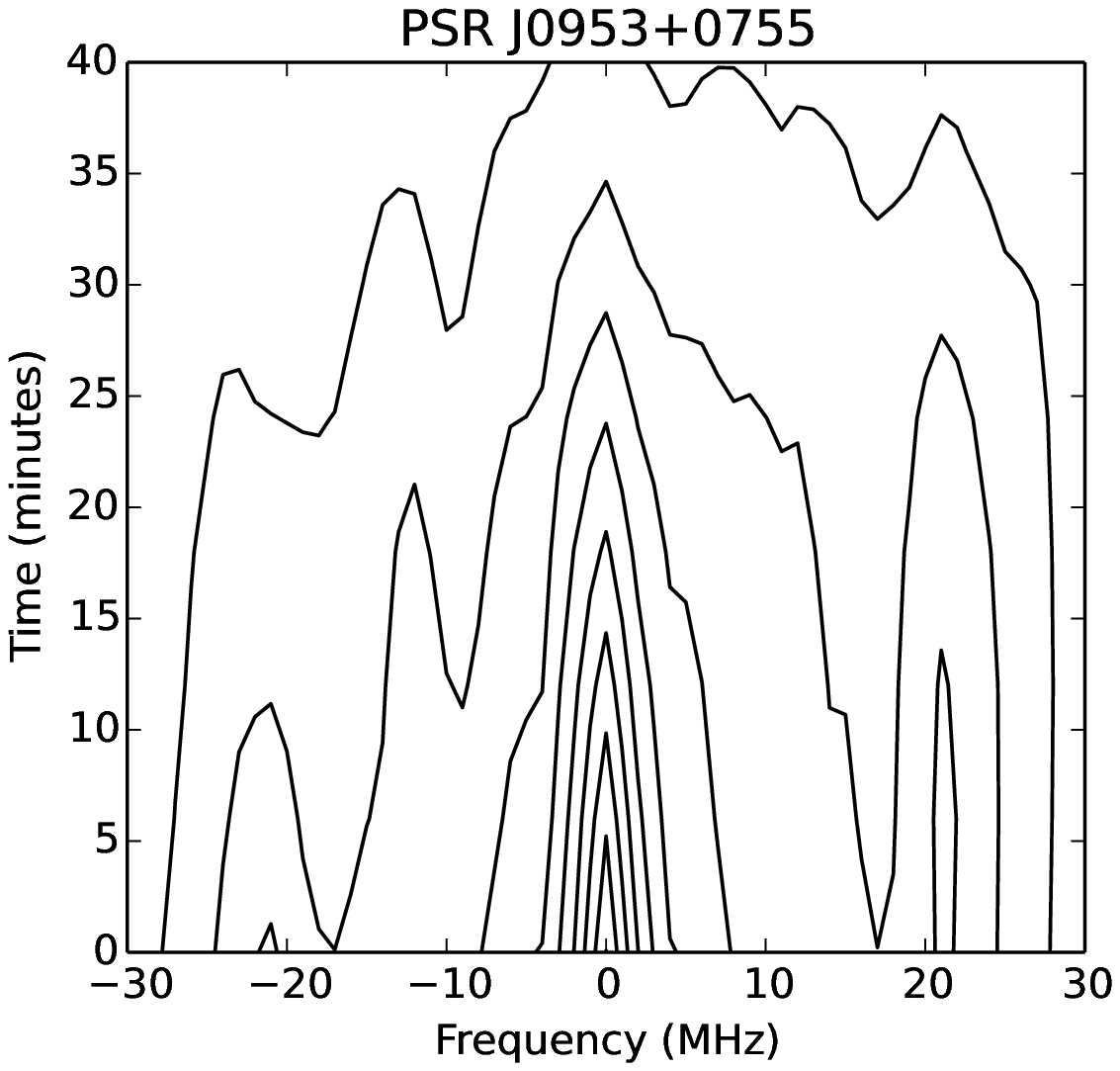}
\includegraphics[scale=0.57]{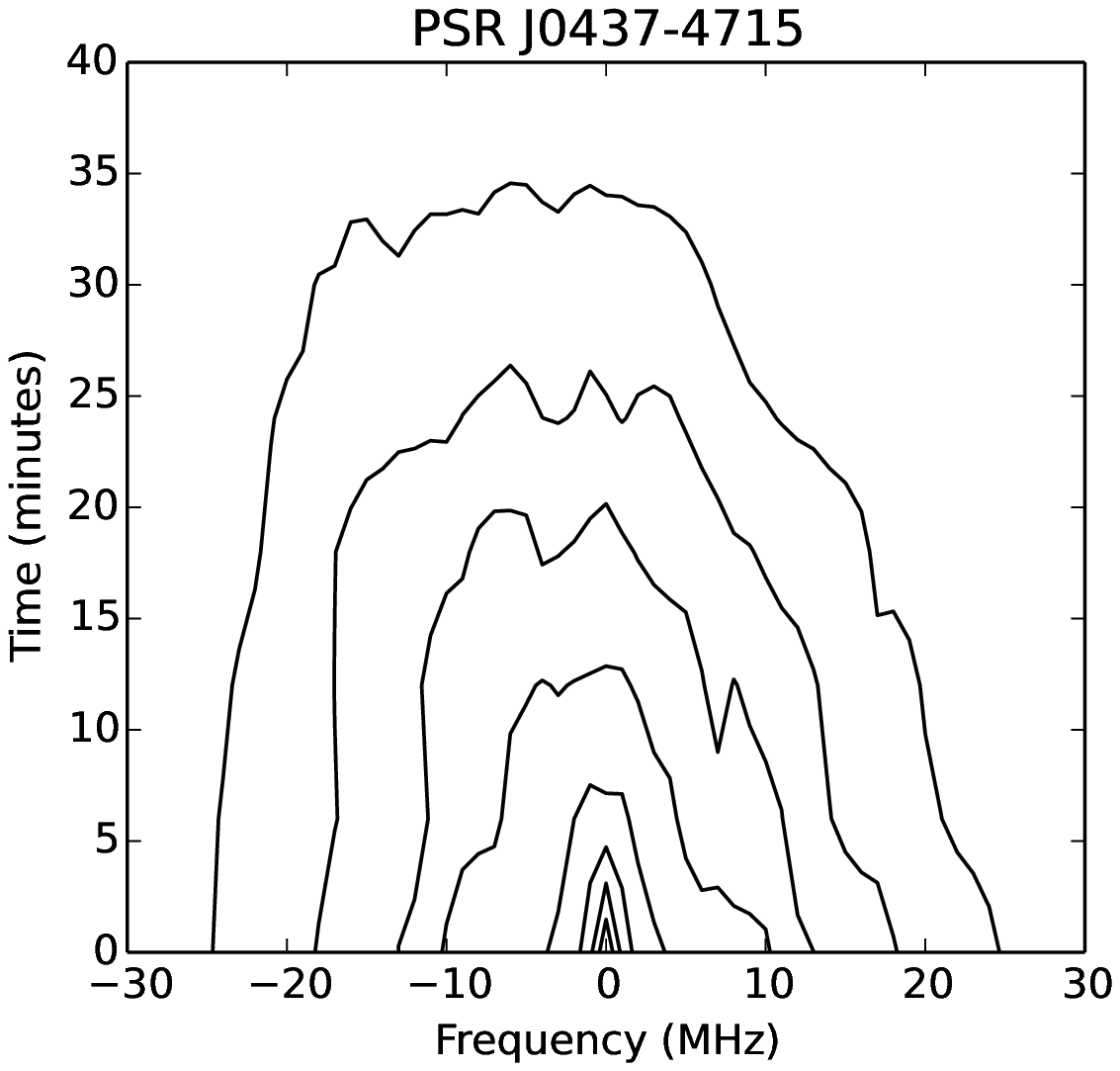}
\caption{Top row: contour plots of the dynamic spectrum for the pulsars PSR~J0953$+$0755 and PSR~J0437$-$4715 during the most extreme levels of variability. The contours for these plots assume that the observations are continuous in time. For PSR~J0953+0755 the contours run from 10 to 50~Jy in steps of 1~Jy; for PSR~J0437-4715 they run from 2 to 8~Jy in steps of 0.25~Jy.  Bottom row: two-dimensional covariance functions of the dynamic spectra for the corresponding pulsars. Ten contour levels are shown that cover the peak to the minimum of the covariance function.}
\label{dynamic_spectrum}
\end{figure*}

Due to the high signal-to-noise ratio of the pulsar detection during this time, we were able to examine the frequency structure of the variability in the image plane. We took the data from the night of 2015-04-14 and for each observation we imaged the data in 30$\times$0.97~MHz sub-bands. For each of the sub-bands in each of the time slots we used the {\sc aegean} forced fit algorithm (discussed above) to fit the flux density at the location of the pulsar. The dynamic spectrum resulting from these measurements is plotted in Figure~\ref{dynamic_spectrum}. We note that our observations are not continuous i.e. each of the snapshot observations integrate for 112 seconds and then return four minutes later to that declination. Figure \ref{dynamic_spectrum} shows four distinct events with discrete time and frequency structure. The peak of the variability seen in some of the sub-bands was even higher than in the full-bandwidth data: a peak flux density of 48.6~Jy was observed at 142~MHz. 

The modulation index in both frequency and time for all measurements in Figure \ref{dynamic_spectrum} (left) gives 85.6\%. As discussed in \cite{Narayan_92} we may expect a modulation index of up to 100\% for diffractive strong scintillation. In some of the frequency and times bins shown in Figure~\ref{dynamic_spectrum} the pulsar is undetected. The result of these non-detections would be to decrease the modulation index as the flux density measurements are only upper limits.  

We followed the method described in \cite{Cordes_86} to calculate a scintillation bandwidth and timescale based on our observations. We calculated the 2D autocorrelation function of the dynamic spectrum which is shown in Figure \ref{dynamic_spectrum} (bottom row). To parameterize the autocorrelation function we fitted a 1D Gaussian in the time and frequency axes, respectively. We followed the definition in \cite{Cordes_86} whereby the half-width half-maximum in the frequency direction defines the scintillation bandwidth. We used the half width at the $1/e$ point to calculate the scintillation timescale. From this analysis we find a scintillation bandwidth of $\Delta \nu_{d}=4.1$~MHz and a scintillation timescale of $\Delta \tau_{d}=28.8$~minutes. We note that due to our broad bandwidth (30~MHz) we expect approximately a factor of three difference in scintillation bandwidth between to the top and the bottom of our band. For all the calculations above we use the central frequency for all scalings. 

We scaled the scintillation bandwidth and timescale reported by \cite{P_C_Nat_1992} under the assumption that $\Delta \nu_{d} \propto \nu^{4.4} $ and $\Delta \tau_{d} \propto \nu^{1.2}$ \citep{Cordes_86}. We find that at 154~MHz the predicted scintillation timescale is 21.6 minutes and the scintillation bandwidth is 4.5~MHz. The predicted scintillation timescale is slightly shorter than our result of 28.8 minutes. The predicted and measured scintillation bandwidths are in good agreement. The amplitude of variability is extreme, but such cases have been reported before e.g. \cite{Galama_97}. 

We can calculate the expected timescale for refractive scintillation ($\tau_{r}$) using the scintillation bandwidth ($\Delta \tau_{d}$) and timescale ($\Delta \nu_{d}$) via the following expression from \cite{Stinebring_1990}:


\begin{equation}
\tau_r = \frac{4}{\pi} \left(\frac{\nu \Delta \tau_d}{\Delta \nu_d}\right) \propto \nu ^{-11/5}.
\label{ref_eq}
\end{equation}

\noindent Using our diffractive scintillation parameters we find  $\tau_{r} = 20.9$ hours at 154 MHz. This is consistent with \cite{gupta_93} who measure a refractive timescale for this pulsar of 3.4 days at 74 MHz (also see \citealt{cole_1970}). There also appears to be a timescale of several hundred days in the Gupta et al. data, the origin of which is unclear but could be inhomogeneties in the ISM.

This pulsar was recently observed with the LWA at 39.4~MHz and a number of giant pulses were detected that had a signal-to-noise ratio greater than 10 times that of the mean pulses \citep{Tsai_2015}. These giant pulses where however typically reported to be rare with approximately 5 per hour (or 0.035\% of the total number of pulse periods). \cite{singal_2012} observed this pulsar at 103~MHz with only 1.6~MHz of bandwidth. Scaling our results to their frequency, the scintillation bandwidth should be 0.7~MHz and the scintillation timescale should be 18 minutes. At certain epochs, \cite{singal_2012} report very strong pulses over the course of 30 minutes. Although they interpret this as giant pulse emission, we consider it much more likely to be the effects of scintillation. Giant pulses are largely broadband in nature \citep{Tsai_2016}, yet we see significant frequency structure in our observations. We therefore conclude that the extreme variability observed in PSR~J0953$+$0755 is consistent with diffractive scintillation and is not intrinsic to the pulsar.  

\subsection{PSR J0437$-$4715}
With a spin period of 5.76~ms and a DM of 2.65~cm$^{-3}$~pc, PSR J0437-4715 is one of the closest and brightest milli-second pulsars \citep{Johnston_1993}. This pulsar is located 7.3 degrees away from the bright (452~Jy at 160~MHz; \citealt{culgoora_95}) double lobed radio galaxy Pictor~A, making this field challenging to image at low frequencies. The main issues arise when Pictor~A is outside of the MWA field of view and is not de-convolved or {\sc clean}ed. This causes side-lobe flux to be scattered across the image, which in turn affects the image fidelity and the quality of the flux scale correction we are able to achieve and apply within that region. For the light-curve shown in Figure~\ref{pulsar_lightcurves} we removed 14 observations that had extreme gain corrections and bad image fidelity. 

We found a modulation index of $M=44.9\%$ with $\chi_{r}^{2}=28.1$. 
\cite{bhat_2014} have studied PSR~J0437$-$4715 using the MWA at 192.6~MHz. For approximately one hour's worth of data with 
20~s~time resolution and 0.64~MHz of frequency resolution, the authors measure the scintillation 
properties. They report a scintillation bandwidth of $\Delta \nu_{d}=1.7$~MHz and a scintillation 
timescale of $\Delta \tau_{d}=4.5$~minutes. Using the frequency scaling $\Delta \nu_{d} \propto \nu^{3.9} $ and time scaling $\Delta \tau_{d} \propto \nu^{1.2}$ (see \citealt{bhat_2014}) these values become  $\Delta \nu_{d}=0.7$~MHz and $\Delta \tau_{d}=3.5$~minutes at 154~MHz. 

We repeated the same analysis described in Section~\ref{J0953} for the night of 2014-10-19, where 
the pulsar had the highest signal-to-noise ratio (a total of six observations). The dynamic spectrum 
is shown in Figure \ref{dynamic_spectrum} and this pulsar is clearly detected in the higher 
frequency resolution images with a flux density peaking around 7~Jy. The dynamic spectrum is much 
more discrete in frequency and time when compared with PSR~J0953$+$0755. From the 2D autocorrelation 
analysis (see Figure \ref{dynamic_spectrum}, bottom row) we find a scintillation bandwidth of 
$\Delta \nu_{d}=3.1$~MHz and scintillation timescale $\Delta \tau_{d}=3.7$~minutes. The scintillation 
timescale of 3.7 minutes from this study is in good agreement with the scaled value of 3.5 minutes 
from \cite{bhat_2014}. The scintillation bandwidth of $\Delta \nu_{d}=3.1$~MHz is however much 
broader than the scaled value of 0.7~MHz found by \cite{bhat_2014}.   

For this pulsar we are only barely resolving the scintles in the frequency direction. 
Potentially this broader value of $\Delta \nu_{d}$ is a result of this under 
sampling and uncertainties in obtaining a meaningful Gaussian fit to the autocorrelation function. 
We conclude that this variability is attributed to diffractive scintillation but note that our scintillation measurements are 
at the lowest frequency to date, and greater frequency resolution would be beneficial in characterising the scintillation further.  

\subsection{PSR J0630$-$2834 (B0628-28)}
The pulsar PSR J0630$-$2834 (\citealt{Large_Nat}) became brighter peaking at $1.2$~Jy for one of 
the observing runs on 2013-12-06 (see Figure~\ref{pulsar_lightcurves}). The flux density then dropped to 
around $0.4$~Jy in the observations six months later. We measure a modulation index of $M=30.0\%$ with a 
$\chi^{2}_{r}=5.8$. This pulsar is at a DM of 34.5~cm$^{-3}$~pc \citep{Hobbs_2004}. 

Scaling the scintillation bandwidth and timescale reported by \cite{Cordes_86}, 
for this pulsar yields $\Delta \nu_{d}=2.2$~kHz and $\Delta \tau_{d}=1.2$~minutes. 
The predicted scintillation bandwidth (from \citealt{Cordes_86}) is almost three orders of magnitude less than our sub-band frequency resolution (0.97~MHz).
We therefore conclude that the variability is not a consequence of diffractive scintillation. 

Using Equation \ref{ref_eq} we find that the refractive scintillation timescale is 74.2 days. The major jump in flux density corresponds to 239 days (about 7 months). The variability seen is this pulsar is more consistent with refractive scintillation with regards to timescale. Averaging all the flux density measurements per night of observing and re-calculating the modulation index yields 24.0\%, which is slightly lower than the 30\% calculated from including all values independently. These values are consistent with the modulation that would be expected from refractive scintillation (see also \citealt{Bhat_99b}). 

\subsection{PSR~J0034$-$0721 (B0031-07)}
PSR J0034$-$0721 was located at the edge of our survey region so the only data available were where the pulsar was $9-15^{\circ}$ from the pointing centre of the observations. In this region the primary beam correction is less accurate. A number of observations were also removed due to excited ionospheric conditions that affected source positions. The DM for this pulsar is 11.4~cm$^{-3}$~pc \citep{Hobbs_2004} and with the usable observations we detect mildly significant variability. We find a modulation index of $M=45.0\%$ with $\chi_{r}^{2}=2.0$. Scintillation bandwidth and timescale values from \cite{Johnston_0034} scaled to 154~MHz are $\Delta \nu_{d}=0.04$~MHz and $\Delta \tau_{d}>8.7$~minutes. 

The expected scintillation bandwidth is much smaller than our sub-band frequency resolution, therefore the variability is unlikely to be caused by diffractive scintillation. Using Equation \ref{ref_eq} we find a refractive timescale of 30 days. The time difference between the final two epochs in Figure~\ref{pulsar_lightcurves}, where the majority of the variability is concentrated, is 13~days. Averaging the flux density measurements per night of observing and calculating the modulation index yields 32.2\% which is lower than for all measurements independently.       

PSR J0034$-$0721 has been shown to undergo nulling (\citealt{0034_1970}; \citealt{Biggs_1992}). The nulls occur for a duration of up to one minute and repeat sudo-randomly every 100 pulses \citep{0034_1970}. Noting that null duration is similar to the length of our observations (112~s), it is plausible that nulling could reduce the flux density significantly in a given observation. We conclude, however, that the cadence of the nulling (every 100 pulses, or every 94 seconds) would not cause the larger modulated, longer term variability seen in our observations (around epoch 70 onwards). Owing to the lower significance of variability ($\chi_{r}^{2}=2.0$) and difficult ionospheric conditions during observing, it is difficult to draw conclusions about the cause of variability for this pulsar, but refractive scintillation seems the most plausible.

\subsection{PSR J0835$-$4510 (B0833-45)}  
\label{0835}
We measure a modulation index of $M=10.7\%$, which is very close to the the average modulation index of two nearby sources $\overline{M}=9.8$\%. This source had a $\chi^{2}_{r}$ = 1.5 meaning it is considered non-variable by our definition (see Section \ref{results}). We do however include it in this discussion as there are some noteworthy features. 

PSR~J0835$-$4510 is a pulsar with spin period 0.09~s and DM of 68.0 cm$^{-3}$~pc. Historical low frequency measurements of this pulsar by the Culgoora Circular Array (CCA; \citealt{culgoora_95}) report flux densities of $S_{80} = 12$~Jy at 80 MHz and $S_{160} = 9$~Jy at 160~MHz with a spectral index of $\alpha=-0.42$. Here we report a mean flux density of $S_{154} = 5.4 \pm 1.6$ which is significantly lower than the archival measurements. There is a distinct turnover in the spectrum (see Figure \ref{spectrum}) which is potentially attributed to pulse broadening due to interstellar scattering \citep{Higgins_71}. 

Differences in flux density between our measurements and \cite{culgoora_95} could be attributed to 
instrumental differences. The CCA consisted of a circular 3~km baseline array and it lacked sensitivity
 to large, diffuse structure. With many short baselines, the MWA is sensitive to both diffuse and point-like emission.
PSR~J0835$-$4510 is embedded in a region of complex morphology, which includes both the pulsar and the Vela supernova remnant. We would therefore expect with its respective spatial sensitivity that the MWA would measure a greater flux density at the location of the pulsar, when compared with the CCA. 

The size of the restoring beam was used to constrain the Gaussian fits to this object. This applies the assumption that this pulsar is represented by a single point source, which is unresolved. Separating the intrinsic flux of the pulsar from the contribution from the supernova remnant is difficult. We tested fitting this source with an unconstrained Gaussian and the reported major and minor axis of that fit were slightly larger than the restoring beam, indicating that this source is slightly resolved. Clearly it is difficult within this region to obtain an accurate flux density via the method we have chosen. 
 
This $\chi^{2}_{r}$ is driven up by the apparent dip in the light-curve around epoch 65, or 2015-01-20. So far we have no explanation for a physical mechanism that would cause this dip but conclude that it is most likely a combination of source fitting errors (discussed above) and difficulty in achieving adequate flux scale correction in such a complex region of the Galactic plane.  

\subsection{Non-variable pulsars}
The remaining pulsars in our sample remained non-variable with $\chi^{2}_{r}<2.0$ and modulation indices comparable to the neighbouring sources. The non-variable pulsars are: PSR~J1057$-$5226; PSR~J1359$-$6038; PSR~J1400$-$6325; PSR~J1453$-$6413; PSR~J1456$-$6843; PSR~J1534$-$5334; PSR~J1651$-$4246; PSR~J1707$-$4053; PSR~J1752$-$2806; PSR~J1820$-$0427; PSR~J1900$-$2600 and PSR~J2048$-$1616. See Table~\ref{pulsar_table} for full details of the statistics. See Figures \ref{pulsar_lightcurves}, \ref{pulsar_lightcurves2} and \ref{pulsar_lightcurves3} for light curves.  

Visual inspection of SUMSS images for the regions around PSR~J1707$-$4053 and PSR~J1400$-$6325 show low-levels of diffuse emission from supernova remnants. For this work this emission is largely unresolved, but we note that a component of the flux density reported for these pulsars may originate from the supernova remnants.   

\subsection{Spectral properties of detected pulsars}
\label{spec_section}
We calculated a spectral energy distribution for each pulsar using an average flux density measurement for all data points from this work, and available data in the literature. A least squares linear regression was used to find the spectral index and error (see Table \ref{pulsar_spectra}). The pulsars PSR~J1453$-$6413, PSR~J1400$-$6325 and PSR~J1534$-$5334 lacked sufficient archival data to calculate spectral indices. 

\begin{figure}
\centering
\includegraphics[scale=0.53]{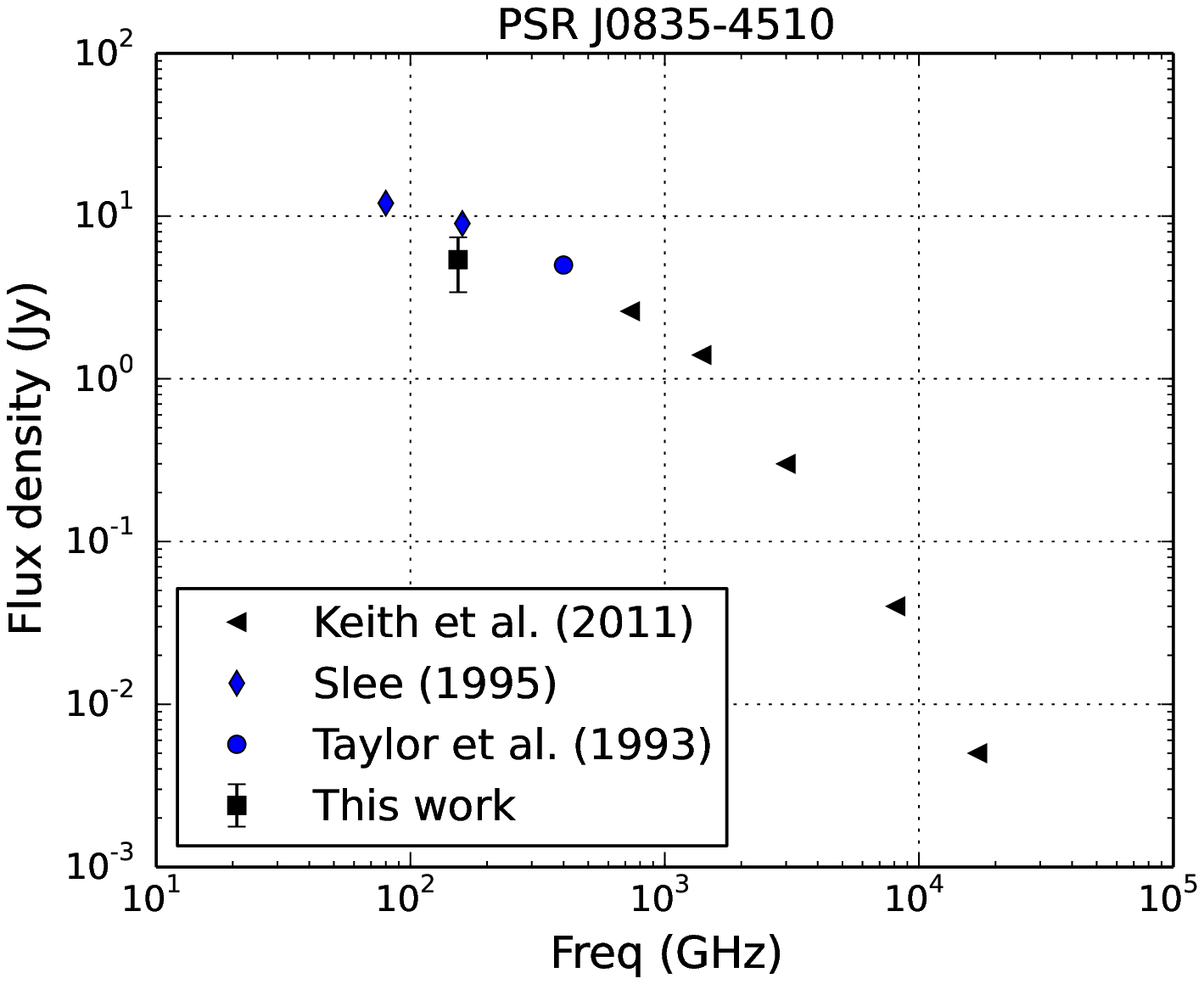}
\includegraphics[scale=0.53]{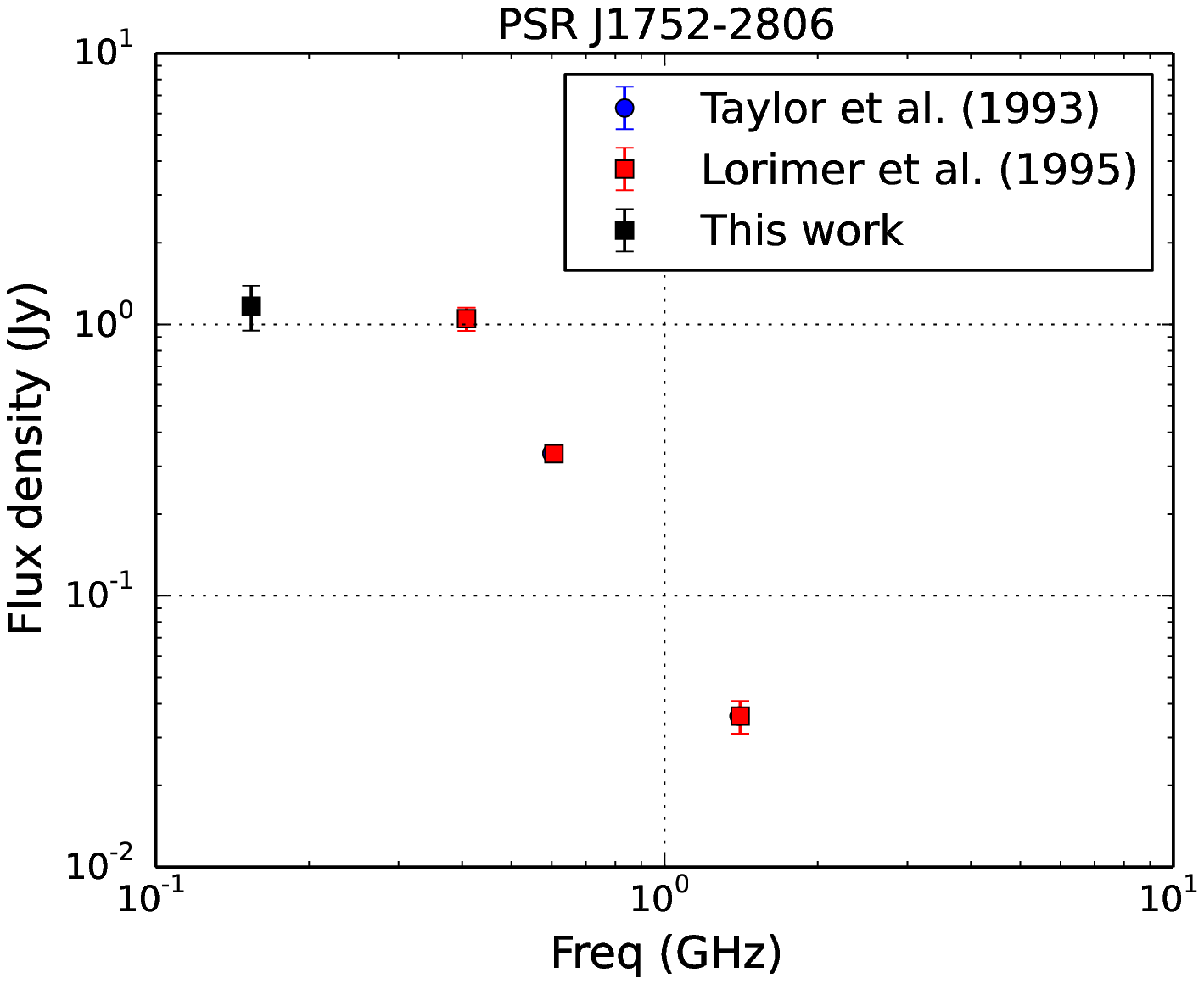}
\includegraphics[scale=0.53]{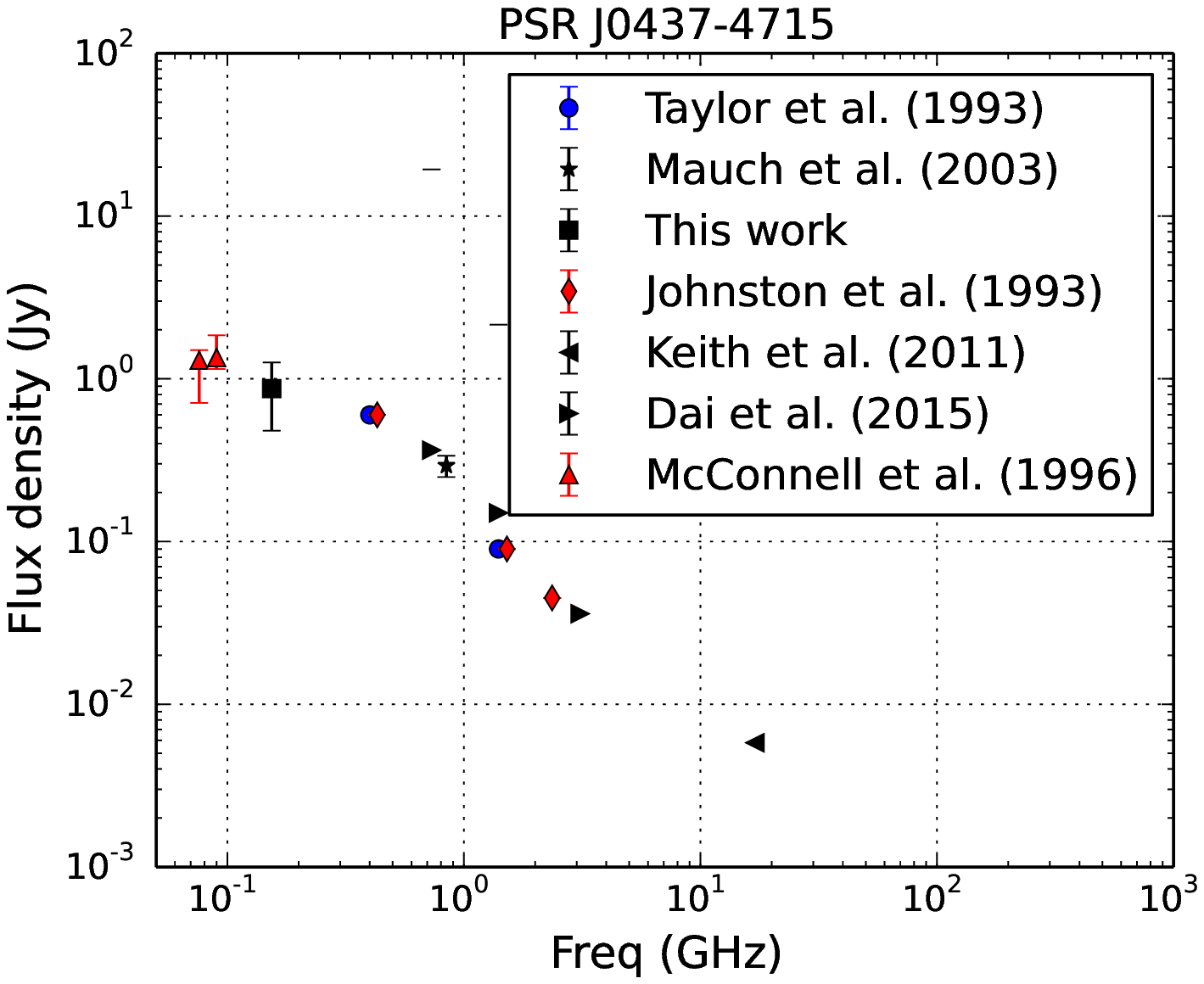}
\caption{Spectral energy distributions of the pulsars PSR~J0835$-$4510, PSR~J1752$-$2806 and PSR~J0437$-$4715. The archival data points are taken from \citealt{Johnston_1993}, \citealt{taylor_93}, \citealt{lorimer_95}, \citealt{culgoora_95}, \citealt{McConnell_96}, \citealt{SUMSS}, \citealt{Keith_2011} and \citealt{Dai_2015}.}
\label{spectrum}
\end{figure} 

In Figure~\ref{spectrum} we show spectra for the pulsars PSR~J0835$-$4510, PSR~J1752$-$2806 and PSR~J0437$-$4715. These pulsars, especially PSR~J1752$-$2806, show significant spectral curvature. We remind the reader of the discussion in Section \ref{0835} regarding the difficulties in obtaining an adequate flux density measurement for PSR~J0835$-$4510. In Figure \ref{spectrum} (left) our data point lies below archival measurements. In the case of PSR~J1752$-2806$, even taking into account a 30\% uncertainty in our flux density scale, our data point is approximately an order of magnitude lower than what would be predicted based on the archival data points of \cite{lorimer_95}. 

The mechanism for this curvature and possible turn over is currently uncertain. Previous studies claim that the abundance of pulsars with low-frequency turnovers is at most 10\% (\citealt{kijak_2011}; \citealt{bates_2013}). Assuming three of the pulsars in our sample of 14 show spectral curvature, this equates to 21\%. This is supported by the recent work of \cite{kuniyoshi_2015}, who show that in a sample of millisecond pulsars, $26\%$ display evidence for turnovers. Results from \cite{Bilous_2015} also support this argument. 

The average of our spectral index values for the pulsars is $\langle\alpha\rangle =-1.5 \pm 0.4$. A broad scatter is potentially a result of uncertainties in our absolute flux scale. This number is however in agreement with \cite{bates_2013} who report an average spectral index of $\langle\alpha\rangle=-1.41 \pm 0.96$, but slightly lower than \cite{maron} who report $\langle\alpha\rangle =-1.8 \pm 0.2$. Our value is also in agreement with \cite{Bilous_2015} who use low frequency measurements and report $\langle\alpha\rangle=-1.4$. \cite{bates_2013} use population synthesis techniques and a likelihood analysis to model the underlying distribution, whereas \cite{maron} derive their value empirically using measurements above 100~MHz only.  

Calculating the mean spectral index of a population of pulsars is dependent on sufficient radio data spanning both MHz and GHz frequencies. It is compounded by frequency dependent selection effects associated with such measurements. Including data below 100~MHz, where the spectral turnover is thought to most commonly occur, results in a flattening of the average spectral index (see \citealt{Malofeev_2000} and \citealt{Bilous_2015}). The MWA has surveyed the Southern sky with frequency coverage between 72 $-$ 231~MHz (see \citealt{GLEAM}), and will contribute to exploring the low-frequency turn over of pulsar spectra further.

\begin{table*}
\centering
\caption{Spectral indices of pulsars calculated using the average flux density from these observations plus arrival data. $\dagger$ Denotes that the pulsar is poorly fit by a power law with significant spectral curvature. The spectral index distributions for these pulsars are shown in Figure~\ref{spectrum}. The pulsar PSR~J1057$-$5226 only had two measurements; therefore we do not report any errors. The references column lists the archival surveys used to calculate the spectral indices. The abbreviations indicate the first author of the survey and year of publication. The full references including the frequencies are in the table footnote.}
\begin{tabular}{|l|r|r|}
\hline
\multicolumn{1}{l}{Pulsar name} & Spectral Index & References  \\
\hline
PSR~J0953$+$0755 & $-1.3 \pm 0.1 $ & L95, T93, S95, C98, D93, C07, M00  \\
PSR~J0437$-$4715$^{\dagger}$ & $ -1.0 \pm 0.1 $ & J93, T93, M03, K11, D15 \\
PSR~J0630$-$2834 & $-1.6 \pm 0.1 $ & L95, T93, C98, D96, C07 \\
PSR~J0034$-$0721 & $-1.6 \pm 0.2 $ & T93, C98, M00 \\
\hline
PSR~J0835$-$4510$^{\dagger}$ & $ -1.3 \pm 0.2 $ & T93  \\
PSR~J1057$-$5226 & $-0.95$ & T93   \\ 
PSR~J1359$-$6038 &  $-1.9 \pm 0.1 $ & T93, N09, M78 \\
PSR~J1456$-$6843 & $-1.1 \pm0.1$ & T93 \\
PSR~J1651$-$4246 & $-2.1 \pm 0.1$ & T93, M78 \\
PSR~J1707$-$4053 & $-2.1 \pm 0.1$ & T93  \\
PSR~J1752$-$2806$^{\dagger}$  &  $ -1.7 \pm 0.4 $ & L95, T93 \\
PSR~J1820$-$0427 & $-2.1 \pm 0.1$ & L95, T93 \\
PSR~J1900$-$2600 & $-1.5 \pm 0.1 $ & L95, T93, C98  \\
PSR~J2048$-$1616 & $-1.7 \pm 0.2$ & L95, T93, N09, C98  \\
\hline
\label{pulsar_spectra}
\end{tabular}
\begin{flushleft}
\noindent
\textit{M78 - \cite{manchester_78} at 408~MHz; J93 - \cite{Johnston_1993} at 430, 1520 and 2360 MHz; T93 - \cite{taylor_93} at 400, 600 and 1400~MHz; L95 - \cite{lorimer_95} at 408, 606, 925, 1408 MHz; S95 - \cite{culgoora_95} at 160~MHz; D96 - \cite{douglas_1996} at 365~MHz; C98 - \cite{NVSS} at 1400~MHz; M03 - \cite{SUMSS} and \cite{SUMSS_murphy} at 843~MHz, M00 - \cite{Malofeev_2000} at 102.5~MHz, C07 - \cite{VLSS} at 74~MHz; N09 - \cite{noutsos_2009} at 1400~MHz and \cite{Dai_2015} at 730, 1400 and 3100 MHz.} 
\end{flushleft}
\end{table*}

\section{Discussion}
\begin{figure*}
\centering
\includegraphics[scale=0.65]{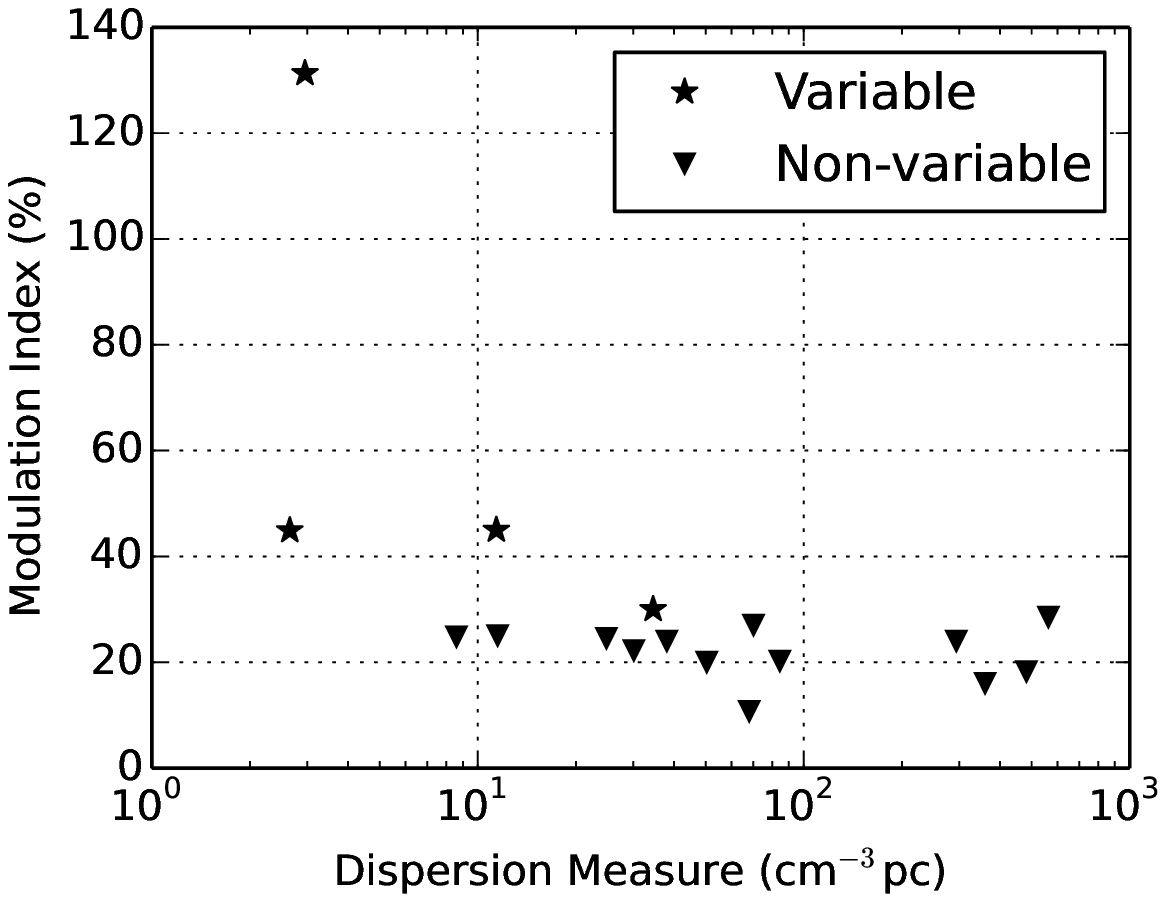}
\includegraphics[scale=0.65]{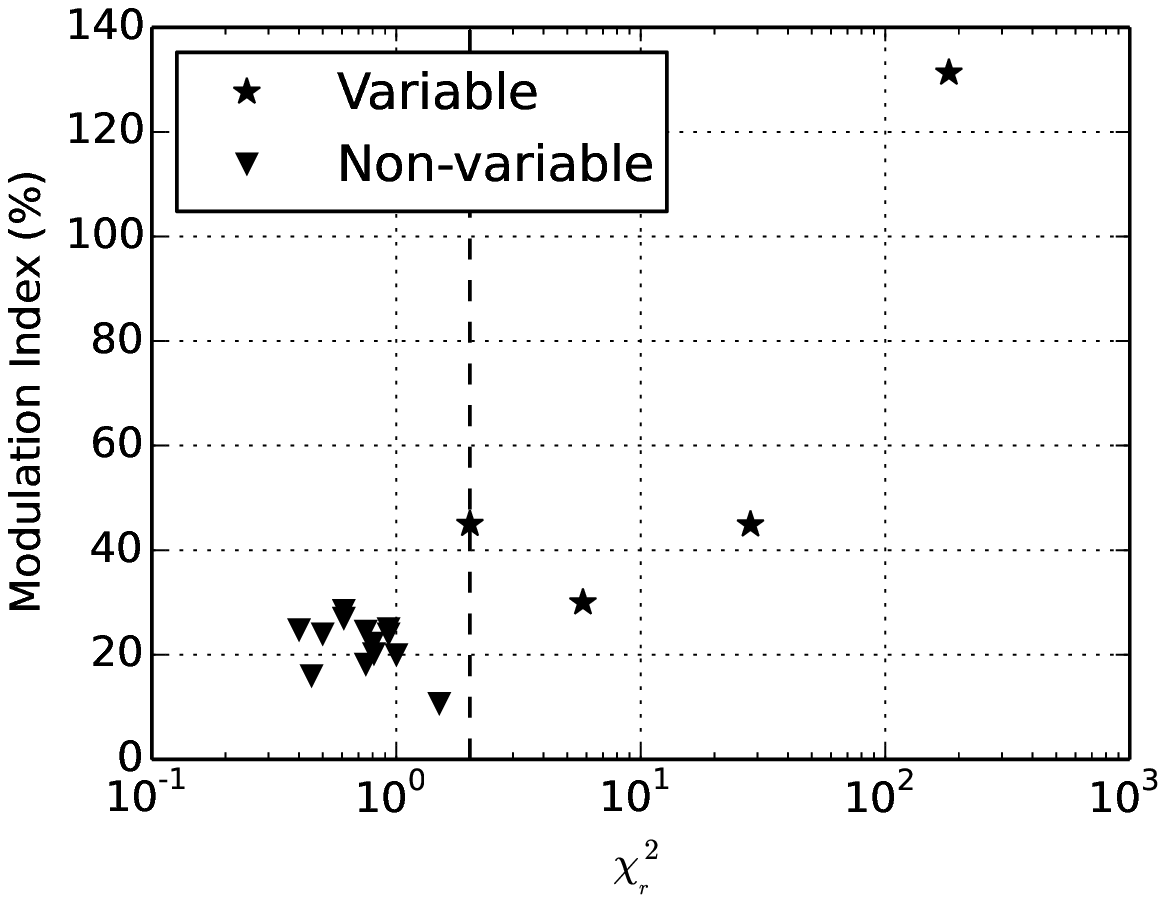}
\caption{Left: plot of dispersion measure versus the modulation index for all pulsars in this sample. The pulsars shown with a starred symbol denote that significant variability was detected ($\chi^{2}_{r}>2$). Right: $\chi_{r}^{2}$ versus modulation index. The dashed line shows $\chi^{2}_{r}=2$.}
\label{DM_vs_M}
\end{figure*}

Figure~\ref{DM_vs_M} shows the dispersion measure versus modulation index (left) and $\chi_{r}^{2}$ versus modulation index (right) for this sample of pulsars. Four pulsars out of our sample of 17 show significant variability. For two of these pulsars (PSR~J0953$+$0755 and PSR~J0437$-$4715) we conclude that the variability is consistent with diffractive scintillation. A further two pulsars (PSR~J0630$-$2834 and PSR~J0034$-$0721) show variability that is best explained by refractive scintillation. This conclusion is less definitive for PSR~J0034$-$0721. 

Two of the pulsars, PSR J2048$-$1616 and PSR J1456$-$6843, show no significant variability despite their low dispersion measure (DM$<15$ cm$^{-3}$ pc). The lack of detection in these pulsars may be related to the probability of sampling bright diffractive scintillation events. This survey is limited by the conservative constraints we place on measurement errors. Reducing these uncertainties may indeed reveal significant variability for these pulsars in future analyses.   

One question we would like to answer is whether we can make new detections of previously unknown pulsars blindly with this method using the MWA, or in the future with the Square Kilometre Array (SKA; \citealt{ska_low})? 
We have shown that MWATS can detect pulsars as transient sources through
their scintillation properties. However, the bandwidth and time averaging
that we perform implies that only those pulsars which have a scintillation 
bandwidth of at least a few MHz at 154~MHz are seen as transients.
Using the \cite{Cordes_Lazio_2002} electron density model we can infer a 
DM and hence distance we could probe with this limit on the scintillation 
bandwidth. This yields a limit of 15.6~cm$^{-3}$~pc or 
a distance of 0.6~kpc (assuming $l=0$, $b=0$). We also need to ensure that the pulsar is
above the detection threshold of MWATS ($100$~mJy). In principle
therefore we could detect a 10~mJy pulsar if the scintillation boost
was a factor of 10 (similar to that seen for PSR~J0953$+$0755).
How many such pulsars exist in our Galaxy?

We simulate a pulsar population using
PsrPopPy\footnote{https://github.com/samb8s/PsrPopPy} (Bates et al. 2014),
drawing spin periods and positions from distributions described by 
Lorimer et al. (2006) and luminosities from a log-normal distribution
(Faucher-Giguere \& Kaspi 2006). DMs were assigned by using the NE2001 model for 
the Galactic distribution of free electrons and the true distances to 
simulated sources. We populate the Galaxy with a population of 
$\sim$130,000 pulsars beaming along our line of sight. Tallying only 
sources with DM$< 15.6$~pc~cm$^{-3}$ and a flux density greater than 10~mJy,
we find $125 \pm 12$ detectable pulsars in our simulations.
The current pulsar catalogue contains some 50 pulsars which obey these
criteria, thus there are of order 75 pulsars yet to be discovered that are within our survey parameters.  

In principle we could probe a much larger volume of the Galaxy for pulsars
if the data could be processed in 1~MHz channels rather than over the entire
32~MHz bandwidth. In this case, although the noise in each image would be
higher, we would be sensitive to much narrower scintillation bandwidths
corresponding to larger distances, increasing the likelihood of finding
pulsars not currently detected by conventional searches.

\section{Conclusion}
With the MWA we have detected significant variability in four pulsars using a sample of only 17 over almost the entire Southern Hemisphere. One of the pulsars (PSR~J0953+0755) shows extreme variability, of order a factor of 60. Both diffractive and refractive interstellar scintillation appear to explain the variability seen in our variable pulsar sample.  

Signal-to-noise and good characterisation of instrumental errors is required to generate adequate variability statistics. Improving upon our current techniques could offer further detections. Continued observations also harbour the possibility of detecting rare and bright events, such as that displayed in PSR~J0953+0755. Future observations with an upgraded MWA with more tiles will allow for further exploration of the pulsar variability parameter space. This also includes refining the flux density measurements of the large number of low signal-to-noise ($3 \sigma$) ratio pulsars found via this work.  

We predict that there are of order 75 pulsars that have not yet been detected via previous high time resolution surveys that could be detected by this method. These pulsars could potentially be of exotic or unusual type. Imaging observations with low frequency widefield interferometers therefore offer a new technique to explore and expand an already diverse population.    

Prospects of exploring diffractive and refractive scintillation in imaging observations with the SKA are intriguing, especially exploring further DM ranges using the increased sensitivity and bandwidth capabilities. The possibility of detecting new pulsars via this imaging method is also promising. Assuming that a number of static continuum and time-domain surveys are completed with SKA then we could contemplate these pulsar surveys being completed commensally. This is true for the data presented in this paper which has been the result of a broad science case blind transient survey (MWATS).     

\section{Acknowledgements}
This scientific work makes use of the Murchison Radio-astronomy Observatory, operated by CSIRO. We acknowledge the Wajarri Yamatji people as the traditional owners of the Observatory site. Support for the operation of the MWA is provided by the Australian Government Department of Industry and Science and Department of Education (National Collaborative Research Infrastructure Strategy: NCRIS), under a contract to Curtin University administered by Astronomy Australia Limited. We acknowledge the iVEC Petabyte Data Store and the Initiative in Innovative Computing and the CUDA Center for Excellence sponsored by NVIDIA at Harvard University. JKS is supported from NSF Physics Frontier Center award number 1430284. DLK and SDC acknowledge support from the US National Science Foundation (grant AST-1412421). Parts of this research were conducted by the Australian Research Council Centre of Excellence for All-sky Astrophysics (CAASTRO), through project number CE110001020. This work was supported by the Flagship Allocation Scheme of the NCI National Facility at the ANU.

\appendix

\label{lastpage}

\end{document}